CLNS 94-1270, CLEO 94-5

# Exclusive Hadronic $B$ Decays to Charm and Charmonium Final States

M.S. Alam,[1] I.J. Kim,[1] B. Nemati,[1] J.J. O'Neill,[1] H. Severini,[1] C.R. Sun,[1] M.M. Zoeller,[1] G. Crawford,[2] C. M. Daubenmier,[2] R. Fulton,[2] D. Fujino,[2] K.K. Gan,[2] K. Honscheid,[2] H. Kagan,[2] R. Kass,[2] J. Lee,[2] R. Malchow,[2] F. Morrow,[2] Y. Skovpen,[2*] M. Sung,[2] C. White,[2] F. Butler,[3] X. Fu,[3] G. Kalbfleisch,[3] W.R. Ross,[3] P. Skubic,[3] J. Snow,[3] P.L. Wang,[3] M. Wood,[3] D.N. Brown,[4] J.Fast ,[4] R.L. McIlwain,[4] T. Miao,[4] D.H. Miller,[4] M. Modesitt,[4] D. Payne,[4] E.I. Shibata,[4] I.P.J. Shipsey,[4] P.N. Wang,[4] M. Battle,[5] J. Ernst,[5] Y. Kwon,[5] S. Roberts,[5] E.H. Thorndike,[5] C.H. Wang,[5] J. Dominick,[6] M. Lambrecht,[6] S. Sanghera,[6] V. Shelkov,[6] T. Skwarnicki,[6] R. Stroynowski,[6] I. Volobouev,[6] G. Wei,[6] P. Zadorozhny,[6] M. Artuso,[7] M. Goldberg,[7] D. He,[7] N. Horwitz,[7] R. Kennett,[7] R. Mountain,[7] G.C. Moneti,[7] F. Muheim,[7] Y. Mukhin,[7] S. Playfer,[7] Y. Rozen,[7] S. Stone,[7] M. Thulasidas,[7] G. Vasseur,[7] G. Zhu,[7] J. Bartelt,[8] S.E. Csorna,[8] Z. Egyed,[8] V. Jain,[8] K. Kinoshita,[9] K.W. Edwards,[10] M. Ogg,[10] D.I. Britton,[11] E.R.F. Hyatt,[11] D.B. MacFarlane,[11] P.M. Patel,[11] D.S. Akerib,[12] B. Barish,[12] M. Chadha,[12] S. Chan,[12] D.F. Cowen,[12] G. Eigen,[12] J.S. Miller,[12] C. O'Grady,[12] J. Urheim,[12] A.J. Weinstein,[12] D. Acosta,[13] M. Athanas,[13] G. Masek,[13] H.P. Paar,[13] J. Gronberg,[14] R. Kutschke,[14] S. Menary,[14] R.J. Morrison,[14] S. Nakanishi,[14] H.N. Nelson,[14] T.K. Nelson,[14] C. Qiao,[14] J.D. Richman,[14] A. Ryd,[14] H. Tajima,[14] D. Schmidt,[14] D. Sperka,[14] M.S. Witherell,[14] M. Procario,[15] R. Balest,[16] K. Cho,[16] M. Daoudi,[16] W.T. Ford,[16] D.R. Johnson,[16] K. Lingel,[16] M. Lohner,[16] P. Rankin,[16] J.G. Smith,[16] J.P. Alexander,[17] C. Bebek,[17] K. Berkelman,[17] K. Bloom,[17] T.E. Browder,[17†] D.G. Cassel,[17] H.A. Cho,[17] D.M. Coffman,[17] P.S. Drell,[17] R. Ehrlich,[17] M. Garcia-Sciveres,[17] B. Geiser,[17] B. Gittelman,[17] S.W. Gray,[17] D.L. Hartill,[17] B.K. Heltsley,[17] C.D. Jones,[17] S.L. Jones,[17] J. Kandaswamy,[17] N. Katayama,[17] P.C. Kim,[17] D.L. Kreinick,[17] G.S. Ludwig,[17] J. Masui,[17] J. Mevissen,[17] N.B. Mistry,[17] C.R. Ng,[17] E. Nordberg,[17] J.R. Patterson,[17] D. Peterson,[17] D. Riley,[17] S. Salman,[17] M. Sapper,[17] F. Würthwein,[17] P. Avery,[18] A. Freyberger,[18] J. Rodriguez,[18] R. Stephens,[18] S. Yang,[18] J. Yelton,[18] D. Cinabro,[19] S. Henderson,[19] T. Liu,[19] M. Saulnier,[19] R. Wilson,[19] H. Yamamoto,[19] T. Bergfeld,[20] B.I. Eisenstein,[20] G. Gollin,[20] B. Ong,[20] M. Palmer,[20] M. Selen,[20] J. J. Thaler,[20] A.J. Sadoff,[21] R. Ammar,[22] S. Ball,[22] P. Baringer,[22] A. Bean,[22] D. Besson,[22] D. Coppage,[22] N. Copty,[22] R. Davis,[22] N. Hancock,[22] M. Kelly,[22] N. Kwak,[22] H. Lam,[22] Y. Kubota,[23] M. Lattery,[23] J.K. Nelson,[23] S. Patton,[23] D. Perticone,[23] R. Poling,[23] V. Savinov,[23] S. Schrenk,[23] and R. Wang[23]

(CLEO Collaboration)

[1]*State University of New York at Albany, Albany, New York 12222*

[2]*Ohio State University, Columbus, Ohio, 43210*

[3]*University of Oklahoma, Norman, Oklahoma 73019*

[4]*Purdue University, West Lafayette, Indiana 47907*

[5]*University of Rochester, Rochester, New York 14627*

[6]*Southern Methodist University, Dallas, Texas 75275*

[7] *Syracuse University, Syracuse, New York 13244*
[8] *Vanderbilt University, Nashville, Tennessee 37235*
[9] *Virginia Polytechnic Institute and State University, Blacksburg, Virginia, 24061*
[10] *Carleton University, Ottawa, Ontario K1S 5B6 and the Institute of Particle Physics, Canada*
[11] *McGill University, Montréal, Québec H3A 2T8 and the Institute of Particle Physics, Canada*
[12] *California Institute of Technology, Pasadena, California 91125*
[13] *University of California, San Diego, La Jolla, California 92093*
[14] *University of California, Santa Barbara, California 93106*
[15] *Carnegie-Mellon University, Pittsburgh, Pennsylvania 15213*
[16] *University of Colorado, Boulder, Colorado 80309-0390*
[17] *Cornell University, Ithaca, New York 14853*
[18] *University of Florida, Gainesville, Florida 32611*
[19] *Harvard University, Cambridge, Massachusetts 02138*
[20] *University of Illinois, Champaign-Urbana, Illinois, 61801*
[21] *Ithaca College, Ithaca, New York 14850*
[22] *University of Kansas, Lawrence, Kansas 66045*
[23] *University of Minnesota, Minneapolis, Minnesota 55455*

## Abstract

We have fully reconstructed decays of both $\bar{B}^0$ and $B^-$ mesons into final states containing either $D$, $D^*$, $D^{**}$, $\psi$, $\psi'$ or $\chi_{c1}$ mesons. This allows us to obtain new results on many physics topics including branching ratios, tests of the factorization hypothesis, color suppression, resonant substructure, and the $B^- - \bar{B}^0$ mass difference.

Typeset using REVTeX

---

*Permanent address: INP, Novosibirsk, Russia

†Permanent address: University of Hawaii at Manoa

## I. INTRODUCTION

Since $B$ mesons were first fully reconstructed in 1983 by CLEO [1] there have been several papers by CLEO [2], [3] and ARGUS [4], [5], [6] which reported branching ratios for exclusive decay modes of $B$ mesons. We present here new data from the CLEO II detector using a high resolution photon detector and a much larger data sample than has been available previously.

We are particularly interested in two-body hadronic $B$ meson decays, which occur through the Cabibbo favored $b \to c$ transition. In these circumstances the dominant weak decay diagram is the spectator diagram, shown in Fig. 1(a). The virtual $W^-$ materializes into either a $\bar{u}d$ or $\bar{c}s$ pair. This pair becomes one of the final state hadrons while the $c$ quark pairs with the spectator anti-quark to form the other hadron. The Hamiltonian [7] , ignoring hard gluon corrections, is

$$H = \frac{G_F}{\sqrt{2}} V_{cb} \left\{ \left[ (\bar{d}u) + (\bar{s}c) \right] (\bar{c}b) \right\} \tag{1}$$

where $(\bar{q}_i q_j) = \bar{q}_i \gamma_\mu (1 - \gamma_5) q_j$, $G_F$ is the Fermi coupling constant, and $V_{cb}$ is the CKM matrix element.

The spectator diagram is modified by hard gluon exchanges between the initial and final quark lines. The effect of these exchanges can be taken into account by use of the renormalization group. These gluons induce an additional term so that the effective Hamiltonian is comprised of two pieces, the original one now multiplied by a coefficient $c_1(\mu)$ and an additional term multiplied by $c_2(\mu)$:

$$H_{eff} = \frac{G_F}{\sqrt{2}} V_{cb} \left\{ c_1(\mu) \left[ (\bar{d}u) + (\bar{s}c) \right] (\bar{c}b) + c_2(\mu) \left[ (\bar{c}u)(\bar{d}b) + (\bar{c}c)(\bar{s}b) \right] \right\} \tag{2}$$

where the $c_i$ are Wilson coefficients evaluated at the mass scale $\mu$. The Wilson coefficients can be calculated from QCD; however, the calculation of rates is inherently difficult because it is unclear at what scale these coefficients should be evaluated. The usual scale is taken to be $\mu \sim m_b^2$. Defining

$$c_\pm(\mu) = c_1(\mu) \pm c_2(\mu) \tag{3}$$

the leading-log approximation gives [8]

$$c_\pm(\mu) = \left( \frac{\alpha_s(M_W^2)}{\alpha_s(\mu)} \right)^{\frac{-6\gamma_\pm}{(33 - 2n_f)}} \tag{4}$$

where $\gamma_- = -2\gamma_+ = 2$, and $n_f$ is the number of active flavors, five in this case.

The Hamiltonian in Eq. (2) leads to the "color suppressed" diagram shown in Fig. 1(b), which reflects the quark pairings in the term multiplied by the coefficient $c_2(\mu)$. Observation of $B \to \psi X_s$ decays, where $X_s$ is a strange meson, gives experimental evidence for the existence of this diagram. Further information on the size of the color suppressed contribution can be obtained from $\bar{B}^0 \to D^0$ (or $D^{*0}$)$X^0$ transitions, where $X^0$ is a neutral meson containing light quarks. In $B^-$ decays, both types of diagrams are present and can interfere. By comparing the rates for $B^-$ and $\bar{B}^0$ decays, the size and the sign of the color suppressed term can be extracted.

Bjorken has suggested [9] that, in analogy to semileptonic decays, two body decays of $B$ mesons that occur via the external spectator process can be expressed theoretically as the product of two independent hadronic currents, one describing the formation of a charm meson and the other the hadronization of the $\bar{u}d$ (or $\bar{c}s$) system from the virtual $W^-$. Qualitatively, he argues that for a $B$ decay with a large energy release the $\bar{u}d$ pair, which is produced as a color singlet, travels fast enough to leave the interaction region without interfering with the formation of the second hadron. The assumption that the amplitude can be expressed as the product of two hadronic currents is called "factorization" in this paper. Several tests of the factorization hypothesis can be made by comparing semileptonic and hadronic $B$ meson decays.

This paper is structured in the following manner: the data sample, detector and reconstruction procedures are described in sections II and III. Branching ratios are given for $B \to D\pi^-$ and $B \to D\rho^-$ modes in section IV. In section V results on branching ratios, polarizations and final state substructure for $B \to D^*\pi^-$, $B \to D^*\rho^-$ and $B \to D^* a_1^-$ are described. Section VI describes a search for $D^{**}$ production in hadronic $B$ decay. This is followed by section VII on exclusive $B$ decays to charmonium, and section VIII on a search for other color suppressed $B$ decays. A $B^- - \bar{B}^0$ mass difference measurement is described in section IX. The interpretation of these results and comparisons to theoretical predictions are discussed in sections X (factorization tests), XI (spin symmetry tests) and XII (determination of the color suppressed amplitude).

## II. DATA SAMPLE AND SELECTION CRITERIA

### A. Data Sample

The data sample used in this paper was collected with the CLEO II detector at the Cornell Electron Storage Ring (CESR). The integrated luminosity is 0.89 fb$^{-1}$ at the $\Upsilon(4S)$ resonance and 0.41 fb$^{-1}$ at energies just below $B\bar{B}$ threshold, henceforth referred to as the continuum. It is natural to assume equal production of charged and neutral $B$'s since the difference between their masses is very small (see section IX). Then there are a total of $935,000 \pm 10,000 \pm 15,000$ charged and the same number of neutral $B$ mesons in this sample.

### B. Detector

The CLEO II detector [10] is designed to detect both charged and neutral particles with excellent resolution and efficiency. The detector consists of a charged particle tracking system surrounded by a time-of-flight (TOF) scintillation system and an electromagnetic shower detector with 7800 thallium-doped cesium iodide crystals. In the "barrel", defined as the region where the angle of the shower with respect to the beam axis lies between 32° and 135°, the r.m.s. energy resolution is given by $\delta E/E(\%) = 0.35/E^{0.75} + 1.9 - 0.1E$ ($E$ in GeV). In the endcap region, located between 18° and 36° from the beam axis, the r.m.s energy resolution is given by $\delta E/E(\%) = 0.26/E + 2.5$ . The tracking system, time-of-flight scintillators, and calorimeter are installed inside a 1.5 T superconducting solenoidal magnet. Immediately outside the magnet are iron and chambers for muon detection. The momentum resolution of the tracking system is given by $(\delta p/p)^2 = (0.0015p)^2 + (0.005)^2$ ($p$ in GeV/c). Ionization loss information ($dE/dx$), provided by the tracking system, is used to identify charged particles in this analysis. The track must have a $dE/dx$ measurement that differs from that expected for the charged particle hypothesis under consideration by less than $3\sigma$ (where henceforth $\sigma$ denotes the r.m.s. resolution).

Muons are identified by a system of drift tubes interleaved with layers of magnet iron. Electron identification utilizes the specific ionization of the track in the drift chamber, the spatial distribution of the energy in the calorimeter, and the ratio of the cluster energy measured in the calorimeter to the track momentum.

### C. Photon Selection

Photon candidates are selected from showers in the calorimeter barrel that have a minimum energy of 30 MeV, are not matched to a charged particle track from the drift chamber, and have a lateral energy distribution consistent with that expected for photons. In the calorimeter endcap the same criteria are applied but the minimum energy requirement is increased to 50 MeV. A small angular region between $32^0$ and $36^0$ degrees in the barrel-endcap overlap region is excluded. Neutral pion candidates are selected from pairs of photons with an invariant mass within $2.5\sigma$ of the known $\pi^0$ mass. These candidates are kinematically fitted with a $\pi^0$ mass constraint.

Candidate $\eta$ mesons are reconstructed in the $\eta \to \gamma\gamma$ mode. They are required to have an invariant mass within 30 MeV of the known $\eta$ mass (547.5 MeV) [14]. The candidates which pass the requirements described above, are kinematically constrained to the $\eta$ mass. Candidate $\eta'$ mesons are reconstructed in the $\eta\pi^+\pi^-$ channel with $\eta \to \gamma\gamma$. Candidate $\omega$ mesons are reconstructed in the $\omega \to \pi^+\pi^-\pi^0$ channel.

### D. Charm meson selection

We select $D^0$, $D^+$, $D^{*+}$ and $D^{*0}$ mesons based on the following criteria. Candidate $D^0$ mesons are identified in the decay modes $D^0 \to K^-\pi^+$, $D^0 \to K^-\pi^+\pi^0$ and $D^0 \to K^-\pi^+\pi^+\pi^-$. Candidate $D^+$ mesons are selected using the $D^+ \to K^-\pi^+\pi^+$ mode. The decay modes, branching ratios and r.m.s. mass resolutions, $\sigma_{m_D}$, are listed in Table I. We use the CLEO [11] absolute branching ratio for $D^0 \to K^-\pi^+$ decays [13], and the Mark III value for $D^+ \to K^-\pi^+\pi^+$ [12]. We use the Particle Data Group (PDG) values [14] for the ratios $\mathcal{B}(D^0 \to K^-\pi^+\pi^0)/\mathcal{B}(D^0 \to K^-\pi^+)$ (where henceforth $\mathcal{B}$ denotes the branching ratio) and $\mathcal{B}(D^0 \to K^-\pi^+\pi^+\pi^-)/\mathcal{B}(D^0 \to K^-\pi^+)$.

Charged $D^*$ candidates are found using the decay $D^{*+} \to \pi^+ D^0$, while neutral $D^*$ candidates are found using the decay $D^{*0} \to \pi^0 D^0$. Other $D^*$ decay modes are not used because they have much poorer signal to background ratios. CLEO branching ratios (Table II) are used for $D^*$ decays [16]. We form $D^{*+}$ and $D^{*0}$ candidates by selecting $D^0$ candidates whose mass is within $2.5\sigma$ of the known $D^0$ mass. Then we require that the $D^*$–$D^0$ mass difference be within $2.5\sigma$ of the measured values [14,17].

### E. Charmonium Meson Selection

We reconstruct the charmonium states $\psi$, $\psi'$ and $\chi_{c1}$, where $\psi$ mesons are selected by their decay into pairs of identified leptons ($e^+e^-$ or $\mu^+\mu^-$ ). We use the MarkIII value [18] $\mathcal{B}(\psi \to l^+l^-) = (5.91 \pm 0.25)\%$ for the $\psi$ to dilepton branching ratio. The kinematics of $B$ decay at the $\Upsilon(4S)$ imply that each lepton in a $\psi$ candidate has a momentum between 0.8 GeV/c and 2.8 GeV/c, with one lepton always having momentum greater than 1.5 GeV/c. For the clean modes $B^- \to \psi K^-$ and $B^0 \to \psi K_S^0$, we obtain good efficiency in the dimuon channel by requiring only one identified muon that penetrates through three interaction lengths. In the dielectron channel, one of the electrons must satisfy a loose electron probability requirement. For modes other than $B^- \to \psi K^-$ and $B^0 \to \psi K_S^0$, both electrons must be identified, or one muon is required to penetrate five interaction lengths and the partner muon is required to penetrate three interaction lengths.

Final state radiation is included in the Monte Carlo simulation of $\psi$ meson decays. For $\psi$'s in the dielectron final states we employ an asymmetric mass cut: $-150 < m(e^+e^-) - m(\psi) <$ 45 MeV in order to reduce the efficiency loss from this source. For the dimuon final state, we require $-45 < m(\mu^+\mu^-) - m(\psi) < 45$ MeV since final state radiation is less significant in this case. (The $\psi$ mass resolution would be 15 MeV (r.m.s.) in the absence of radiation). In these mass windows the efficiency for detecting $\psi$ mesons in the dielectron and the dimuon final states are 48.1% and 67.8% for the looser cuts, and are 45.7% and 42.4% when both leptons are identified.

The decay modes $\psi' \to e^+e^-$, $\psi' \to \mu^+\mu^-$, and $\psi' \to \pi^+\pi^-\psi$, are used to select $\psi'$ candidates. The recontruction of the leptonic decays follows the procedure outlined for $\psi$ mesons. Pion candidates for the third decay mode are required to have dE/dx measurements consistent with the pion hypothesis. In addition, tracks that have been identified as part of a $K_S^0$ decay are rejected. It has been shown that the $\pi^+\pi^-$ invariant mass spectrum from $\psi'$ decays favor larger values relative to that expected from phase space [18]. We require the $\pi^+\pi^-$ invariant mass to be between 0.45 and 0.58 GeV [19]. For $\psi'$ mesons reconstructed through the decay $\psi' \to \pi^+\pi^-\psi$ we require the $\psi' - \psi$ mass difference, $\delta m = m_{\psi'} - m_\psi$, to be between 0.568 and 0.599 GeV. $\chi_{c1}$ mesons are reconstructed by their decay into a photon and a $\psi$ meson. We require the photon be in the good portion of the barrel calorimeter ($|\cos\theta| < 0.71$). If the photon candidate forms an invariant mass within $-5$ to 3 standard deviations of the known $\pi^0$ mass when combined with any other photon in the event, it is rejected.

## III. $B$ MESON RECONSTRUCTION PROCEDURES

### A. Candidate Selection

After selecting $D$, $D^*$ or charmonium candidates we combine them with one or more additional hadrons to form $B$ candidates. The measured sum of charged and neutral energies, $E_{\rm meas}$ of correctly reconstructed $B$ mesons produced at the $\Upsilon(4S)$ must equal the beam energy, $E_{\rm beam}$, to within the experimental resolution. Depending on the $B$ decay mode, $\sigma_{\Delta E}$, the r.m.s. resolution on the energy difference $\Delta E = E_{\rm beam} - E_{\rm meas}$ varies from 8 to 46 MeV. The modes considered and the corresponding $\sigma_{\Delta E}$ values are given in Tables III–VII and Tables IX–X. We divide the $B$ candidates into a signal sample where $\Delta E$ is consistent with zero within 2.5 $\sigma$, and a "sideband" sample consisting of two intervals, one with $\Delta E$ positive and the other negative, both $2.5\sigma$ wide and at least $3\sigma$ away from $\Delta E = 0$. For decay modes with large $\sigma_{\Delta E}$, we restrict the sideband width so that the maximum value of $\Delta E$ is less than one pion mass. This avoids contamination from the $B$ decay mode with an additional pion. These $\Delta E$ sidebands are used to study the background shape.

For $B$ decay modes with a fast $\rho^-$ the energy resolution depends on the momenta of the pions from the $\rho^-$ decay. The momenta of the charged and neutral pions are correlated; a fast $\pi^-$ accompanies a slow $\pi^0$ and vice-versa. This correlation is most conveniently formulated as a function of the helicity angle $\Theta_\rho$, the angle in the $\rho^-$ rest frame between the direction of the $\pi^-$ and the $\rho^-$ direction in the lab frame. When $\cos\Theta_\rho = +1$, the resolution in the energy measurement is dominated by the momentum resolution on the fast $\pi^-$. In contrast, when $\cos\Theta_\rho = -1$, the largest contribution to the energy resolution comes from the calorimeter energy resolution on the fast $\pi^0$. Typically $\sigma_{\Delta E}$ varies linearly between 20 MeV at $\cos\Theta_\rho = -1$ and 40 MeV at $\cos\Theta_\rho = 1$. The energy resolution from a Monte Carlo simulation for one such mode ($B^- \to D^0\rho^-$) is shown in Fig. 2 as a function of the $\rho^-$ helicity angle. The energy difference resolutions for modes containing a $\rho^-$ are given in Tables III–VI.

In addition to the above selection criteria, events are required to satisfy $\mathrm{R}_2 < 0.5$ where $\mathrm{R}_2$ is the ratio of the second Fox-Wolfram moment to the zeroth moment determined using charged tracks and unmatched neutral showers [20]. A sphericity angle cut is applied to further reduce continuum background. The sphericity angle $\Theta_s$ is the angle between the sphericity axis of the particles which form the $B$ candidate and the sphericity axis of the other particles in the event [21]. For a jet-like continuum event, the absolute value of this angle is small; while for a $B\bar{B}$ event, the two axes are almost uncorrelated. Requiring $|\cos\Theta_s| < 0.7$ typically removes about 80% of the continuum background, while retaining 70% of the $B$ decays. The sphericity cut used here depends on the number of pions which accompany the $D$ or $D^*$ meson. For final states with a $D^*$ and a single (2, 3) pion(s) we require $|\cos\Theta_s| < 0.9$ (0.8, 0.7). For all modes which contain a $D$ and a single (2) pion(s) in the final state, we demand that $|\cos\Theta_s| < 0.8$. In modes with $\psi$ mesons, we maximize the efficiency by applying no sphericity angle cut.

To determine the signal yield and display the data we form the beam constrained mass

$$M_B^2 = E_{\rm beam}^2 - \left(\sum_i \vec{p_i}\right)^2, \tag{5}$$

where $\vec{p_i}$ is the momentum of the $i$-th daughter of the $B$ candidate. The resolution in this variable is about 2.7 MeV [22] and is about a factor of ten better than the resolution in invariant mass. The width is dominated by the CESR beam energy spread rather than by detector resolution.

For a specific $B$ decay chain, such as $B^- \to D^0\pi^-, D^0 \to K^-\pi^+\pi^0$, we allow only one candidate per event to appear in the $M_B$ distribution. If there are multiple candidates with $M_B > 5.2$ GeV, the entry with the smallest absolute value of $\Delta E$ is selected.

### B. Background Studies

In order to extract the number of signal events it is crucial to understand the shape of the background in the $M_B$ distributions. There are two contributions to this background, continuum and other $B\bar{B}$ decays. The fraction of background from continuum events varies between about 58% and 91% depending on the $B$ decay mode [23].

We expect that the $M_B$ distribution from the $\Delta E$ sidebands will give a good representation of the background shape. To verify this, a Monte Carlo simulation of $B\bar{B}$ events was used to show that the $\Delta E$ sidebands can be used to accurately model the shape of the $B\bar{B}$ background under the signal in the beam constrained mass distributions (see Fig. 3). In continuum data, the $\Delta E$ sidebands also model the shape of the background in the signal region. The sum of the $B\bar{B}$ Monte Carlo and continuum data agrees in shape with the $\Delta E$ sidebands in data (see Fig. 4). Therefore, the $\Delta E$ sidebands can be used to model the shape of the background under the signal in data. The $M_B$ distributions for $\Delta E$ sidebands in data for several modes are shown in Fig. 5. All of these can be fitted with a linear background below $M_B$=5.282 GeV, and a smooth kinematical cutoff at the endpoint, which we choose to be parabolic. The distributions of $M_B$ for wrong-sign combinations (e.g. $\bar{D}^0\pi^+$), wrong-charge combinations (e.g. $D^+\pi^+$), and continuum data can also be adequately fitted with this functional form (henceforth referred to as the CLEO background shape). To determine the number of signal events from the $M_B$ spectrum in the $\Delta E$ interval centered on zero, we use the background function as determined from the sidebands and a Gaussian signal with a fixed width of 2.64 MeV.

### C. Efficiency Studies

In order to extract branching ratios, detection efficiencies are determined from a Monte Carlo simulation of the CLEO II detector. The accuracy of the simulation is checked in several ways. We select radiative Bhabha events ($e^+e^- \to e^+e^+\gamma$) using only calorimeter information and then embed the tracks into hadronic events. We find that the efficiency for the detection of charged tracks above 225 MeV/c is correct to better than 2%. The Monte Carlo simulation of charged tracks with transverse momenta below 225 MeV/c is more complicated since these tracks do not traverse the entire drift chamber. The accuracy of the simulation is verified using the $D^*$ decay angle distribution of inclusive $D^{*+} \to D^0\pi^+$, $D^0 \to K^-\pi^+$ decays which must be symmetric after efficiency correction. The simulation of low $p_T$ tracks agrees with the Monte Carlo simulation for $100 < p < 225$ MeV/c. However, the efficiency for tracks in this momentum range is known to only $\pm 5\%$.

The accuracy of the photon detection efficiency can be verified by comparing the ratio of branching ratios of $\eta \to \pi^0\pi^0\pi^0$ and $\eta \to \gamma\gamma$ to the average ratio given by the PDG [14]. This test indicates that the single photon efficiency is modelled to better than $\pm 2.5\%$. Other checks of the $\pi^+$ and $\pi^0$ detection efficiency are performed by comparing the yield of fully reconstructed $D^0 \to K^-\pi^+\pi^0$ decays with the yield of partially reconstructed $D^0 \to$ $K^-\pi^0(\pi^+)$ where the $\pi^+$ is not detected. Additional consistency checks have been performed by comparing inclusive $D^{*+}$ and $D^{*0}$ cross sections in the continuum, and by comparing the ratios $\mathcal{B}(D^0 \to K^-\pi^+\pi^-\pi^+)$ $/\mathcal{B}(D^0 \to K^-\pi^+)$ and $\mathcal{B}(\eta \to \pi^-\pi^+\pi^0)/\mathcal{B}(\eta \to \gamma\gamma)$ to the values in the PDG compilation [14].

## IV. BRANCHING RATIOS FOR $D\pi^-$ AND $D\rho^-$ FINAL STATES

We reconstruct the decay modes $\bar{B}^0 \to D^+\pi^-$, $\bar{B}^0 \to D^+\rho^-$, $B^- \to D^0\pi^-$, and $B^- \to D^0\rho^-$ following the procedures described in sections II and III. There is an additional complication for the analysis of the $B \to D\rho^-$ modes. Events which are consistent with the decay chain $B \to D^*\pi^-$, $D^* \to D\pi^0$ have the same final state particles and thus form a potential background. We eliminate this background by discarding events for which the $D^* - D$ mass difference is consistent with the $D^*$ hypothesis. This veto does not reduce the efficiency for $B \to D\rho^-$. A Monte Carlo simulation of $B\bar{B}$ decays shows a broad enhancement in the signal region for $B^- \to D^0\pi^-$ and $B^- \to D^0\rho^-$ (see Fig. 3). This enhancement contains contributions from $B \to D^{*0}X$, $D^{*0} \to D^0\gamma$ transitions which can be modeled with the CLEO background shape.

To select $B \to D\rho^-$ channels we impose additional requirements on the $\pi^-\pi^0$ invariant mass and decay angle. Specifically, we require that $|m(\pi^-\pi^0)-770| < 150$ MeV/c$^2$. Since the decay $B \to D\rho^-$ is fully longitudinally polarized (helicity zero due to angular momentum conservation), a cut on the $\rho$ helicity angle is imposed ($|\cos\Theta_\rho| > 0.4$) [24]. The beam constrained mass distributions for $B \to D\pi^-$ and $B \to D\rho^-$ are shown in Fig. 9. Fig. 6 shows the $\rho$ helicity angle distributions (with the cut on the helicity angle removed) for $\bar{B}^0 \to D^+\rho^-$ and for $B^- \to D^0\rho^-$ after $B$ mass sideband subtraction. After efficiency correction, these distributions are given by the functional form:

$$\frac{dN}{d\cos\Theta_\rho} = \frac{\Gamma_L}{\Gamma}\cos^2\Theta_\rho + 0.5(1 - \frac{\Gamma_L}{\Gamma})\sin^2\Theta_\rho \tag{6}$$

where $\Gamma_L/\Gamma$ is the fraction of longitudinal polarization. The fit gives $\Gamma_L/\Gamma = 1.07 \pm 0.05$ for $B^- \to D^0\rho^-$ and $\Gamma_L/\Gamma = 0.92 \pm 0.07$ for $\bar{B}^0 \to D^+\rho^-$. These results are consistent with full polarization as expected and thus provide a consistency check of the background subtraction and efficiency correction. Monte Carlo simulation shows that most of the $B\bar{B}$ backgrounds in $B \to D\rho^-$ decays are due to combinations with an incorrectly reconstructed low momentum $\pi^0$. Therefore a fit to the beam constrained mass distribution with $\cos\Theta_\rho < -0.4$ is also performed as a consistency check of the analysis [25]. These results agree with the branching ratios obtained using the full range of helicity angle.

The $\pi^-\pi^0$ invariant mass distribution for the $B$ signal region ($\pm 6.5$ MeV of the nominal $B$ mass), is shown in Figs. 7 and 8 after $B$ sideband subtraction. Fitting this distribution to the sum of a Breit Wigner and a parameterization of non-resonant $B \to D\pi^-\pi^0$ decay [26] we find that fewer than 2.5% (at 90% C.L.) of the events in the $B$ mass peak arise from non-resonant decays, after applying the helicity angle cut and restricting the $\pi^-\pi^0$ mass to lie in the rho mass region. The observed $D\pi^-\pi^0$ events are consistent with $B \to D\rho^-$ and any non-resonant contribution can be neglected.

The resulting branching ratios for $B \to D\pi^-$ and $B \to D\rho^-$ are given in Tables III and IV.

Two systematic errors are quoted on the branching ratios. The first includes contributions from background shape ($\sim 5\%$), Monte Carlo statistics ($2-4\%$), and the uncertainty in the modeling of the tracking and $\pi^0$ detection efficiencies (which depend on the multiplicity of the decay mode as described earlier) and the relative $D^0$ branching fractions. The second systematic error contains the errors from the $D^+ \to K^-\pi^+\pi^+$ ($\pm 14\%$) and $D^0 \to K^-\pi^+$ absolute branching ratios ($\pm 2.7\%$).

## V. MEASUREMENTS OF $D^*(N\pi)^-$ FINAL STATES

### A. Branching Ratios

We now consider final states containing a $D^*$ meson and one, two or three pions. These include the $B \to D^*\pi^-$ , $B \to D^*\rho^-$, and $B \to D^*a_1^-$ decay channels. A cut on the $D^*$ helicity angle, $|\cos\theta_{D^*}| > 0.4$, is made for $B \to D^*\pi$ but not for $D^*\rho$ and $D^*a_1$. The beam constrained mass distributions for the $\bar{B}^0 \to D^{*+}\pi^-$ and $B^- \to D^{*0}\pi^-$ are shown in Fig. 10. Our results for the decays $\bar{B}^0 \to D^{*+}\pi^-$ and $B^- \to D^{*0}\pi^-$ are listed in Tables V and VI. The first error quoted on the branching ratios is statistical, followed by two systematic errors. The first systematic error contains contributions from the uncertainties in the efficiency for charged track finding, the uncertainty in photon detection, variations in event yield from changes in background shape, Monte Carlo statistics and the relative $D^0$ branching fractions. The second systematic error contains the errors on the $D^0 \to K^-\pi^+$ and $D^* \to D^0\pi$ branching ratios.

Fig. 10 shows the beam constrained mass distributions for the $\bar{B}^0 \to D^{*+}\rho^-$ and $B^- \to D^{*0}\rho^-$. To study the resonant substructure in $\bar{B}^0 \to D^{*+}\pi^-\pi^0$ the cut on the $\pi^-\pi^0$ mass is removed. For events in the $B$ signal region ($|M_B - 5.280| < 0.006$ GeV) the $\pi^-\pi^0$ spectrum is examined after subtracting the $\pi^-\pi^0$ spectrum from the low $B$ mass sideband ($5.2 < M_B < 5.26$ GeV). The background subtracted $\pi^-\pi^0$ invariant mass spectrum is then fitted to the sum of a Breit Wigner and a polynomial parameterization of non-resonant $\bar{B}^0 \to D^{*+}\pi^-\pi^0$ obtained from a Monte Carlo simulation. Fig. 11 shows the fit to the background subtracted $\pi^-\pi^0$ invariant mass spectrum. The fit gives an upper limit of less than 6 non-resonant $\bar{B}^0 \to D^{*+}\pi^-\pi^0$ events in the $\rho$ mass window at the 90% confidence level. This implies that the non-resonant contribution to the $\bar{B}^0 \to D^{*+}\rho^-$ decay is less than 9% at the 90% confidence level. If we instead take the shape of the non-resonant component from a $D^{**}(2420)\pi^+$ Monte Carlo we obtain a similar limit for the non-$\rho$ component. A similar study has been made of $B^- \to D^{*0}\rho^-$ which also shows a negligible non-resonant component. The branching ratios for $B \to D^*\rho$ can be found in Tables V and VI. In Fig. 12 we show the $M_B$ distributions for $B^- \to D^{*0}\pi^-\pi^-\pi^+$ and $\bar{B}^0 \to D^{*+}\pi^-\pi^-\pi^+$ where the $\pi^-\pi^-\pi^+$ invariant mass is required to be in the interval $1.0 < \pi^-\pi^-\pi^+ < 1.6$ GeV. To show that this signal arises dominantly from $a_1^-$ we also present the $M_B$ distributions for the $a_1$ sidebands $0.6 < \pi^-\pi^-\pi^+ < 0.9$ GeV and $1.7 < \pi^-\pi^-\pi^+ < 2.0$ GeV (Fig. 13), where there are signals of $15 \pm 6$ and $0 \pm 5.5$ events for the $D^{*+}$ and $D^{*0}$ channels respectively. The sideband signals are $18\pm6\%$ ($0\pm13$ %) of the signals in the $a_1$ peak, as compared to the expectation of about 10% from the tails of a Breit-Wigner distribution. In Figs. 14 and 15 we show the $\pi^-\pi^-\pi^+$ mass distributions for a $B \to D^* a_1^-$ Monte Carlo simulation, a $B \to D^*\pi^-\rho^0$ non-resonant background simulation, and the data events in the $B$ signal region with the scaled $B$ mass sideband subtracted. The $a_1$ meson is parameterized in the Monte Carlo simulation as a Breit-Wigner resonance shape with $m_{a_1} = 1182$ MeV and $\Gamma_{a1} = 466$ MeV. The fit gives upper limits of less than 4.2 and 4.6 non-resonant events at the 90% confidence level. This implies that the non-resonant components in this decay are less than 9.4% and 10.6% at the 90 % confidence level. We have verified that a $D^{**}(2420)\rho^-$ Monte Carlo simulation gives a similar limit for the non-$a_1$ component. Our results for $B$ meson decays into final states containing a $D^*$ meson and three charged pions are also listed in Tables V and VI.

### B. Polarization in $B \to D^{*+}\rho^-$ decays

The sample of fully reconstructed $B \to D^{*+}\rho^-$ decays can be used to measure the $D^{*+}$ and $\rho^-$ polarizations. By comparing the measured polarizations in $\bar{B}^0 \to D^{*+}\rho^-$ with the expectation from the corresponding semileptonic $B$ decay a test of the factorization hypothesis can be performed (see section X C).

The polarization is obtained from the helicity angle distribution. The $\rho$ helicity angle $\Theta_\rho$ was defined earlier. The $D^{*+}$ helicity angle $\Theta_{D^*}$ is the angle between the $\pi^+$ direction and $B$ direction in the $D^{*+}$ rest frame.

The momentum in the laboratory for pions from the $D^{*+}$ decay which are emitted in the backward hemisphere ($\cos\Theta_{D^*} < 0$ in our convention) extends from 160 MeV/c down to about 100 MeV/c. In this momentum range, the reconstruction efficiency for charged tracks is reduced and becomes momentum dependent.

Before examining the $\bar{B}^0 \to D^{*+}\rho^-$ decay mode, we perform a consistency check of the efficiency correction and analysis procedure by measuring the polarization in $\bar{B}^0 \to D^{*+}\pi^-$. Since $B$ mesons and pions are pseudoscalars, the $D^{*+}$ mesons from the decay $\bar{B}^0 \to D^{*+}\pi^-$ will be longitudinally polarized giving a $\cos^2\Theta_{D^*}$ distribution. The same procedure used in the analysis of the $\bar{B}^0 \to D^{*+}\rho^-$ polarization is applied to this case. After performing the sideband subtraction and correcting for efficiency, [27] we obtain the $D^{*+}$ helicity angle distribution shown in Fig. 16 (c). A fit to this distribution gives $\Gamma_L/\Gamma = 106 \pm 7\%$ which is consistent with the expectation from angular momentum conservation of $\Gamma_L/\Gamma = 100\%$.

We now proceed to measure the polarization in $\bar{B}^0 \to D^{*+}\rho^-$ decays. After integration over $\chi$, the angle between the normals to the $D^{*+}$ and the $\rho^-$ decay planes, the helicity angle distribution can be expressed as follows [28]:

$$\frac{d^2\Gamma}{d\cos\Theta_{D^*}d\cos\Theta_\rho} \propto \frac{1}{4}\sin^2\Theta_{D^*}\sin^2\Theta_\rho(|H_{+1}|^2 + |H_{-1}|^2) + \cos^2\Theta_{D^*}\cos^2\Theta_\rho|H_0|^2 \qquad (7)$$

The fraction of longitudinal polarization is defined by [28]

$$\frac{\Gamma_L}{\Gamma} = \frac{|H_0|^2}{|H_{+1}|^2 + |H_{-1}|^2 + |H_0|^2} \qquad (8)$$

If longitudinal polarization dominates, both the $D^{*+}$ and the $\rho^-$ helicity angles will follow a $\cos^2\Theta$ distribution, whereas in the case of transverse polarization we will observe a $\sin^2\Theta$ distribution for both helicity angles.

To measure the polarization we combine the helicity angle distributions for the three $D^0$ submodes in the $B$ signal region (defined by $|M_B - 5.280| < 0.006$ GeV) and then subtract the helicity angle distribution of the scaled sideband (defined by $5.200 < M_B < 5.260$ GeV). We fit the resulting helicity angle distributions to the functional form given in equation (6).

From the fit to the $D^{*+}$ helicity angle distribution, we find $\Gamma_L/\Gamma = (85 \pm 8)\%$, and from the corresponding fit to the $\rho$ helicity angle distribution we find $\Gamma_L/\Gamma = (97 \pm 8)\%$. The results of the fits [29] are shown in Fig. 16(a) and (b) . The statistical error can be reduced by taking advantage of the correlation between the two helicity angles (See Fig. 17). The most precise result can be extracted by performing an unbinned two dimensional likelihood fit to the joint $(\cos\Theta_{D^*}, \cos\Theta_\rho)$ distribution. This method gives

$$(\Gamma_L/\Gamma)_{\bar{B^0} \to D^{*+}\rho^-} = (93 \pm 5 \pm 5)\% \tag{9}$$

The systematic error contains the uncertainties due to the background parameterization and the detector acceptance.

## VI. MEASUREMENTS OF $D^{**}$ FINAL STATES

In addition to the production of $D$ and $D^*$ mesons, the charm quark and spectator anti-quark can also hadronize as a $D^{**}$ meson. The $D^{**0}(2460)$ has been observed experimentally and identified as the $J^P = 2^+$ state, while the $D^{**0}(2420)$ has been identified as the $1^+$ state [14]. These states have full widths of approximately 20 MeV. Two other states, a $0^+$ and another $1^+$ are predicted but have not yet been observed. Presumably this is due to their large intrinsic widths. There is evidence for $D^{**}$ production in semileptonic $B$ decays [30,31], and it is possible that the $D^{**}$ can also be seen in hadronic $B$ decays.

In order to search for $D^{**}$ mesons from $B$ decays we first study the final states $B^- \to D^{*+}\pi^-\pi^-$ and $B^- \to D^{*+}\pi^-\pi^-\pi^0$. In the latter case we require that one $\pi^-\pi^0$ invariant mass is consistent with the $\rho^-$ mass. The reactions $B^- \to D^{*+}\pi^-\pi^-$ and $B^- \to D^{*+}\pi^-\pi^-\pi^0$ have not been observed clearly in past experiments [3,6] and are not expected to occur in a simple picture in which the $c$ quark plus spectator antiquark form a $D^*$. We combine the $D^{*+}$ with a $\pi^-$ to form a $D^{**}$ candidate. $D^{**}$ candidates lying within one full width of the nominal mass of either a $D^{**0}(2420)$ or a $D^{**0}(2460)$ are then combined with a $\pi^-$ or $\rho^-$ to form a $B^-$ candidate.

We have also searched for $D^{**}$ production in the channels $D^+\pi^-\pi^-$ and $D^0\pi^-\pi^+$. Since $D^{**0}(2420) \to D\pi$ is forbidden, we only search for $D^{**0}(2460)$ in the $D\pi\pi$ final state. For this subset of modes, we require the $D\pi$ mass to lie within $\pm$ 1.5 $\Gamma$ ($\pm$28 MeV) of the nominal $D^{**}(2460)$ mass.

Figs. 18 and 21 show the $B$ mass distributions for combinations of $D^{**0}(2460)$ or $D^{**0}(2420)$, and $\pi^-$ or $\rho^-$. In the $D^{**0}(2420)\pi^-$ mode, there is an excess of 8.5 events in the $B$ peak region with an estimated background of 1.5 events. The probability that the excess is due to a background fluctuation is $4 \times 10^{-6}$ which indicates that this is a significant signal. In this channel we give the branching ratio in Table VII, while for the other five combinations where the probability that the observed events are the result of a background fluctuation is larger, we give upper limits. Our results are consistent with theoretical predictions [32,33] based on the factorization hypothesis (Table VIII).

We have also investigated the final states $D^{*+}\pi^-\pi^-$ ($D^+\pi^-\pi^-$ and $D^0\pi^+\pi^-$) where the $D^{(*)}$ and a charged pion are not constrained to lie in any particular mass interval. To observe these signals, the background from the final states $D^{(*)}\rho^-$ must be suppressed. Since the $D^{(*)}\rho^-$ final state is highly polarized (see section V B), it is possible that the slow $\pi^0$ from the $\rho^-$ decay can be exchanged for a slow charged pion from the decay of the other $B$ meson. To eliminate this background, we make cuts on the cosines of the helicity angles, $\Theta_{D^*}$ and $\Theta_{D^{**}}$, where $\Theta_{D^{**}}$ is calculated for the $D^{*+}\pi^-_{\rm slow}$ system with $\pi^-_{\rm slow}$ being the slower $\pi^-$ of the two. This helicity angle is defined as the angle between the $B$ and the fast $\pi^-$ in the rest frame of $D^{*+}\pi^-_{\rm slow}$ system. We require $\cos\Theta_{D^{**}} < 0.8$ and $|\cos\Theta_{D^*}| < 0.7$. For the $B \to D\pi\pi$ modes, a similar cut, $\cos\Theta_\rho < 0.7$, is made using the $\pi\pi$ system. In Figs. 20 and 19, we show the $M_B$ candidate mass distributions. There is a significant signal in $B \to D^{*+}\pi^-\pi^-$. For the other two modes we quote upper limits in Table VII.

## VII. EXCLUSIVE $B \to$ CHARMONIUM DECAYS

### A. Introduction

In $B$ decays to charmonium the $c$ quark from the $b$ combines with a $\bar{c}$ quark from the virtual $W^-$ to form a charmonium state. This process is described by the color suppressed diagram shown in Fig. 1(b). By comparing $B$ meson decays to different final states with charmonium mesons the dynamics of this decay mechanism can be investigated. The decay modes $\bar{B}^0 \to \psi K^0$ and $\bar{B}^0 \to \psi' K^0$ are of special interest since the final states are CP eigenstates which can be used to determine one of the three CP violating angles accessible to study in $B$ decays. It is also possible to use the $\bar{B}^0 \to \psi K^{*0}$ decay (where $K^{*0} \to K^0_S\pi^0$) to measure this CP asymmetry. However, this final state has even CP if the orbital angular momentum $L$, between the $\psi$ and $K^{*0}$ is 0 or 2, and odd CP for $L = 1$. If both CP states are present the CP asymmetry will be diluted. A measurement of CP violation in this channel may be possible if one of the CP states dominates, or if a detailed moments analysis of the various decay components is performed [34]. We present a measurement of the polarization in the decay $\bar{B}^0 \to \psi K^{*0}$ which allows us to determine the fractions of the two CP states.

### B. Branching Ratios

$B$ meson candidates are formed by combining a charmonium and a strange meson candidate. We reconstruct $K^0_S$ decays into $\pi^+\pi^-$ pairs which have vertices displaced from the beam axis by greater than 5 mm. Only the $K^+\pi^-$ channel is used to form $K^{*0}$ mesons while $K^{*-}$ candidates are reconstructed in the decay channels $K^0_S\pi^-$ and $K^-\pi^0$. The $K\pi$ invariant mass must be within $\pm 75$ MeV of the nominal $K^*$ mass. In the $B^- \to \psi K^{*-}$, $K^{*-} \to K^-\pi^0$ mode, only the half of the $K^{*-}$ helicity angle distribution with a fast $\pi^0$ is used. The $\psi$ and $\psi'$ candidates decaying into lepton pairs were kinematically constrained to the known mass values in order to improve the resolution on $\Delta E$. Using the procedures described in section III we reconstruct $B$ meson candidates and obtain the beam constrained mass distributions shown in Figs. 22, 23 and 24. The corresponding branching ratios are listed in Table IX.

The systematic errors in the branching ratio measurements include contributions from number of $B$ mesons (2.5%), tracking efficiencies (2% per charged track), $\pi^0$ detection efficiency (5%), $dE/dx$ efficiency (2% per identified track), lepton detection efficiency (2% per lepton), Monte Carlo statistics (1.5 - 6%), the $\psi$ leptonic branching ratio (4.2%) and the branching ratios for $\psi'$ and $\chi_{c1}$ decays.

The results for the $B^-$ and $B^0$ decay modes can be combined using isospin symmetry to determine the vector to pseudoscalar production ratio

$$\frac{\mathcal{B}(B \to \psi K^*)}{\mathcal{B}(B \to \psi K)} = 1.71 \tag{10}$$

The revised Bauer-Stech-Wirbel (BSW) model [8] predicts a value of 1.61 for this quantity. This model uses the ratio of $B \to K^*/B \to K$ form factors determined from harmonic oscillator wavefunctions and assumes that the factorization hypothesis is valid for internal spectator decays.

### C. Polarization in $\psi K^*$

After integration over the azimuthal angle between the $\psi$ and the $K^*$ decay planes, the angular distribution in $B \to \psi K^*$ decays can be written as [28]

$$\frac{d^2\Gamma}{d\cos\Theta_\psi d\cos\Theta_{K^*}} \propto \frac{1}{4}\sin^2\Theta_{K^*}(1+\cos^2\Theta_\psi)(|H_{+1}|^2+|H_{-1}|^2)+\cos^2\Theta_{K^*}\sin^2\Theta_\psi|H_0|^2, \tag{11}$$

where the $K^*$ helicity angle $\Theta_{K^*}$ is the angle between the kaon direction in the $K^*$ rest frame and the $K^*$ direction in the $B$ rest frame, $\Theta_\psi$ is the corresponding $\psi$ helicity angle, and the $H_{\pm 1,0}$ are the helicity amplitudes.

There are 29 $\bar{B}^0 \to \psi K^{*0}$ candidates and 13 $B^- \to \psi K^{*-}$ candidates. After correcting for detector acceptance, we perform an unbinned maximum likelihood fit to the double differential distribution described in the equation above. The fit gives the fraction of longitudinal polarization in $B \to \psi K^*$ as

$$\left(\frac{\Gamma_L}{\Gamma}\right)_{B\to\psi K^*} = 0.80 \pm 0.08 \pm 0.05 \tag{12}$$

The systematic error in this measurement is dominated by the uncertainty in the acceptance. The efficiency corrected distributions for each of the helicity angles $\cos\Theta_\psi$ and $\cos\Theta_{K^*}$ are shown in Fig. 25.

This result can be compared to the theoretical predictions of Kramer and Palmer [66] which depends on the unmeasured $B \to K^*$ form factor. Using the BSW model to estimate the form factor, they find $\Gamma_L/\Gamma = 0.57$. Using Heavy Quark Effective Theory (HQET) and experimental measurements of the $D \to K^*$ form factor, they obtain $\Gamma_L/\Gamma = 0.73$.

The decay mode $B \to \psi K^*$ may not be completely polarized, but it is dominated by a single CP eigenstate ($CP = -1$ produced with $L = 1$). This mode will therefore be useful for measurements of CP violation.

## VIII. SEARCH FOR COLOR SUPPRESSED DECAYS

We search for $B$ decays which can occur via an internal $W$-emission graph, but which do not lead to final states with charmonium [39]. One expects that these decays will be suppressed relative to decays which occur via the external $W$-emission graph. For the internal graph the colors of the quarks from the virtual $W$ must match the colors of the $c$ quark and the accompanying spectator antiquark. In a simple picture, one expects that the suppression factor should be about 1/18 for decays involving $\pi^0$, $\rho^0$ and $\omega$ mesons [36], but in heavy quark decays the effects of gluons cannot be neglected. These decays can be used to test QCD based calculations [8] which predict suppression factors of order 1/50. If color-suppressed $B$ decay modes are not greatly suppressed then these modes could be useful for CP violation studies [38].

We search for color-suppressed decay modes of $B$ mesons which contain a single $D$ meson (or $D^*$ meson) in the final state. The relevant color-suppressed modes are given in Table X. We use the decay modes $\eta \to \gamma\gamma$, $\omega \to \pi^+\pi^-\pi^0$ and $\eta^{'} \to \eta\pi^+\pi^-$, followed by $\eta \to \gamma\gamma$ [37].

For decays of a pseudoscalar meson into a final state containing a pseudoscalar and a vector meson, a helicity angle cut of $|\cos\Theta_V| > 0.4$ is used [40]. No convincing signals were found in the decay modes that were examined. Upper limits on the branching ratios for color-suppressed modes are given in Table X. The 90% confidence level upper limits are calculated using the prescription described by the PDG [41]. These upper limits take into account the systematic uncertainty in the background level as well as the systematic uncertainty in the detection efficiency. In Figs. 26, 27, and 28 we show the fitted distributions for the color-suppressed modes with the fit superimposed on each plot. Upper limits on the ratios of color-suppressed modes to normalization modes are given in Table XI.

## IX. THE $B^-$ – $\bar{B}^0$ MASS DIFFERENCE

We now proceed to measurements of the $\bar{B}^0$ and $B^-$ masses and the mass difference between them. For this analysis we use the decays $B^- \to \psi K^-$, $\bar{B}^0 \to \psi K^{*0}$, $B^- \to D^0\pi^-$, $B^- \to D^0\rho^-$, $B^- \to D^{*0}\pi^-$, $B^- \to D^{*0}\rho^-$, $\bar{B}^0 \to D^+\pi^-$, $\bar{B}^0 \to D^+\rho^-$, $\bar{B}^0 \to D^{*+}\pi^-$, and $\bar{B}^0 \to D^{*+}\rho^-$ for which the signal to background ratio is large. For the decays $B^- \to D^0\pi^-$ and $B^- \to D^0\rho^-$ only the $D^0 \to K^-\pi^+$ mode is used. The $M_B$ distributions for the sum of these modes are shown in Fig. 29. We have a total of 362 $B^-$ and 340 $\bar{B}^0$ signal events. The data are fitted with a Gaussian of fixed width (2.7 MeV) determined by Monte Carlo simulation. The width is assumed to be the same for all modes. The fitted masses for each mode and their statistical errors are given in Table XII. We apply a correction for initial state radiation as described in Ref. [42], of magnitude -1.1±0.5 MeV, to arrive at the final values for the $B^-$ and $\bar{B}^0$ masses of $5278.8 \pm 0.2 \pm 0.5 \pm 2.0$ MeV and $5279.2 \pm 0.2 \pm 0.5 \pm 2.0$ MeV, respectively. The first systematic error results from the uncertainty in the initial state radiation correction. The second systematic error is due to the uncertainty in the absolute value of the CESR energy scale, which is determined by calibrating to the known $\Upsilon$(1S) mass [43].

The mass difference is determined to be 0.41±0.25±0.19 MeV. This is more accurate than the masses themselves because the beam energy uncertainty cancels, as do many systematic errors associated with the measurement errors on the charged tracks and $\pi^0$ mesons. The remaining systematic error is found by making a number of tests of the stability of our result.

A systematic shift of 0.12 MeV is produced by using different background shapes for the $B^-$ and $\bar{B}^0$ modes [44]. We have also investigated the effect of changing the photon energy calibration. A change of 0.5%, the quoted systematic error, results in a 0.15 MeV change in the fitted $B$ mass in both the $D^{*+}\rho^-$ and $D^{*0}\rho^-$ final states. This effect almost completely cancels in the mass difference measurement where it contributes an error of <0.03 MeV. This is because the shift of the $\pi^0$ energy in the $\rho^-$ cancels in the difference leaving only the shift of the energy of the slow $\pi^0$ from the $D^{*0}$ which is uncorrelated with the direction of the $B$ meson. A similar test where we scale the measured momentum of the slow pion from the $D^{*+}$ by 5 %, also does not affect the mass difference for the same reason.

We have also checked the stability of the result with changes in event samples. For example, we have used only half of the $\cos\Theta_\rho$ distributions in $B \to D^{(*)}\rho$ modes and we have used less stringent lepton identification criteria for the $B \to \psi K^-$ mode. We estimate a systematic of 0.15 MeV from these studies.

The different sources of systematic errors are listed in Table XIII and are combined in quadrature. We compare our result with previous results in Table XIV.

There are several models which predict the isospin mass difference to be between 1.2 and 2.3 MeV which are larger than the value reported here [45]. However, Goity and Hou ($-0.5 \pm 0.6$ MeV) and Lebed (0.89 MeV) can accomodate this small mass difference in their models [46].

## X. TESTS OF THE FACTORIZATION HYPOTHESIS

### A. Introduction

Our large data sample has made possible the precise branching ratio and polarization measurements discussed above. In the following sections we address many important questions about non-leptonic $B$ meson decay.

By comparing rates and polarizations of semileptonic and hadronic decays we can perform tests of the factorization hypothesis, which is the basis of most theoretical treatment of hadronic $B$ decays. In analogy to semileptonic decays, where the amplitude factorizes into the product of a leptonic and hadronic current since leptons are not sensitive to the strong interaction, it is possible that two body decays of $B$ mesons which occur via the external spectator process may be expressed theoretically as the product of two independent hadronic currents, one describing the formation of a charm meson and the other the hadronization of the $\bar{u}d$ (or $\bar{c}s$) system from the virtual $W^-$.

There are few models of hadronic $B$ decays. Those which exist predict widths of two-body decays and assume the validity of the factorization hypothesis. Although factorization fails in many $D$ decays [47], it is hoped that factorization will be a better approximation in $B$ decays due to the larger energy release present [48].

If factorization is valid, then heavy quark effective theory, henceforth referred to as HQET [49], could provide a reliable, model independent framework for the calculation of properties of non-leptonic $B$ meson decays. In addition, if factorization holds, then measurements of non-leptonic $B$ decays may be used to extract fundamental parameters of the Standard Model. For instance the CKM matrix element $V_{ub}$ can be determined from $B^0 \to \pi^+\pi^-$ or $\bar{B}^0 \to D_s^-\pi^+$, and the decay constant $f_{D_s}$ can be determined from $\bar{B}^0 \to D_s^- D^{*+}$.

### B. Branching Ratio Tests

Assuming factorization, the effective Hamiltonian Eq. (2) for a non-leptonic $B$ decay can be written as a product of two hadronic currents. Consider the case of $\bar{B}^0 \to D^{*+}h^-$, where h is a hadron. The amplitude for this reaction is

$$A = G_F/\sqrt{2}\ V_{cb}V_{ud}^*\langle h^-(p)|(\bar{d}u)|0\rangle\langle D^{*+}|(\bar{c}b)|\bar{B}^0\rangle \tag{13}$$

where $V_{ud}$ is the well measured CKM factor from the $W^- \to \bar{u}d$ vertex. The first hadron current, which creates the $h^-$ from the vacuum, is related to the decay constant $f_h$, and is known for $h = \pi, \rho$. We have

$$\langle h^-(p)|(\bar{d}u)|0\rangle = -if_h p_\mu, \tag{14}$$

where $p_\mu$ is the $h^-$ four momentum. The other hadron current can be found from semileptonic $\bar{B}^0 \to D^{*+}\ell^-\bar{\nu}_\ell$ decays. Here the amplitude is the product of a lepton current and the hadron current that we seek to insert in Eq. (13). Factorization can be tested experimentally by verifying whether the relation

$$\frac{\Gamma\left(\bar{B}^0 \to D^{*+}h^-\right)}{\frac{d\Gamma}{dq^2}\left(\bar{B}^0 \to D^{*+}l^-\bar{\nu}_l\right)\Big|_{q^2=m_h^2}} = 6\pi^2 c_1^2 f_h^2 |V_{ud}|^2, \tag{15}$$

is satisfied, where $q^2$ is the four momentum transfer from the $B$ meson to the $D^*$ meson. Since $q^2$ is also the mass of the lepton-neutrino system, by setting $q^2 = m_h^2$ we are simply requiring that the lepton-neutrino system has the same kinematic properties as the $h^-$ in the hadronic decay. The $c_1^2$ term accounts for hard gluon corrections. Here we use $c_1 = 1.1 \pm 0.1$ as deduced from perturbative QCD. The error in $c_1$ reflects the uncertainty in the mass scale at which the coefficent $c_1$ should be evaluated [50]. For the case where $h^- = \pi^-$ and $c_1$=1, equation 15 was found to be satisfied by Bortoletto and Stone [51]. In the following the left hand side of Eq. (15) will be denoted $R_{\rm exp}$ and the right hand side will be denoted $R_{\rm Theo}$.

This type of factorization test can also be performed using $\bar{B}^0 \to D^{*+}h^-$ decays where $h^- = \rho^-$ or $a_1(1260)^-$. For the $\rho^-$ case Eq. (15) becomes:

$$R = \frac{\Gamma(\bar{B}^0 \to D^{*+}\rho^-)}{\frac{d\Gamma}{dq^2}(B \to D^* l\ \nu)|_{q^2=m_\rho^2}} = 6\pi^2 c_1^2 f_\rho^2 |V_{ud}|^2 \tag{16}$$

where the semileptonic decay is evaluated at $q^2 = m_\rho^2 = 0.60$ GeV$^2$. The decay constant on the right hand side of this equation can be determined from $e^+e^- \to \rho^0$ or from $\tau$ decays. The first method leads to $f_\rho = 215 \pm 4$ MeV. Taking into account the $\rho$ width, Pham and Vu [52] find that $\Gamma(\tau^- \to \nu\rho^-) = \ 0.804 G_F^2/16\pi\ |V_{ud}^2| M_\tau^3 f_\rho^2$ which gives f$_\rho$ = 212.0 $\pm$ 5.3

MeV [53]. We take the first value. We also perform this test for $\bar{B}^0 \to D^{*+} a_1^-$ where we use $f_{a_1} = 205 \pm 16$ MeV [33]. To derive numerical predictions for branching ratios, we must interpolate the observed differential $q^2$ distribution [54] for $\bar{B} \to D^* \ell\ \nu$ to $q^2 = m_\pi^2$, $m_\rho^2$, and $m_{a_1}^2$, respectively. Until this distribution is measured more precisely theoretical models must be used for the slope of the distribution. Thus the results are stated below for different models. Fortunately, the spread in the theoretical models which describe $\bar{B} \to D^* \ell\ \nu$ is small (see Fig. 30).

We now have all the required ingredients [55] for the test with decay rates (see Table XV). Using the extrapolation of the $q^2$ spectrum [56] from the WSB model as the central value, we obtain from Eqs. (15) and (16) the results given in Table XVI.

If we form ratios of branching fractions some of the systematic uncertainties on $\mathrm{R}_{\mathrm{exp}}$ will cancel, as does the QCD correction $c_1$ in $\mathrm{R}_{\mathrm{theor}}$. For example in the case of $D^{*+}\rho^-/D^{*+}\pi^-$, the expectation from factorization is given by $\mathrm{R}_{\mathrm{theor}}(\rho)/\mathrm{R}_{\mathrm{theor}}(\pi)$ times the ratio of the semileptonic branching ratios evaluated at the appropriate $q^2$ values. In Table XVII we show the comparison of the data, the expectation from factorization as defined above and two theoretical predictions of Bauer, Stech and Wirbel (BSW) [64], and Reader and Isgur (RI) [33].

From the measurements described above, we find that at the present level of precision, there is agreement between experiment and the expectation from factorization for the $q^2$ range: $0 < q^2 < m_{a_1}^2$.

### C. Factorization and Angular Correlations

More subtle tests of the factorization hypothesis can be performed by examing the polarization in $B$ meson decays into two vector mesons. This idea was suggested by Körner and Goldstein [57]. Again, the underlying principle is that hadronic decays are analogous to the appropriate semileptonic decays evaluated at a fixed value of $q^2$. For instance, the ratio of longitudinal to transverse polarization $(\Gamma_L/\Gamma_T)$ in $\bar{B}^0 \to D^{*+}\rho^-$ should be equal to the corresponding ratio for $B \to D^* l^-\ \bar{\nu}$ evaluated at $q^2 = m_\rho{}^2 = 0.6$ GeV$^2$.

$$\frac{\Gamma_L}{\Gamma_T}(\bar{B}^0 \to D^{*+}\rho^-) = \frac{\Gamma_L}{\Gamma_T}(B \to D^* l^- \bar{\nu_L})|_{q^2=m_\rho^2} \tag{17}$$

The advantage of this method is that it is not affected by QCD corrections [58].

For $B \to D^*\ l\ \nu$ decay, longitudinal polarization dominates at low $q^2$. Near $q^2 = q^2_{\mathrm{max}}$, by contrast, transverse polarization dominates. There is a simple physical argument for the behaviour of the form factors near these two kinematic limits. Near $q^2 = q^2_{\mathrm{max}}$, the $D^*$ is almost at rest. Its small velocity is uncorrelated with the $D^*$ spin, so all three possible $D^*$ helicities are equally likely. As $q^2 \to q^2_{\mathrm{max}}$ we expect $\Gamma_T/\Gamma_L = 2$. At $q^2 = 0$, the $D^*$ has the maximum possible momentum, while the lepton and neutrino are collinear and travel in the direction opposite to the $D^*$ with their helicities aligned to give $S_z = 0$. Thus, near $q^2 = 0$ longitudinal polarization is dominant.

For $\bar{B}^0 \to D^{*+}\rho^-$, Rosner predicts 88% longitudinal polarization from the argument described above [59]. Similar results can be extracted from the work of Neubert [60] and Kramer et al. [68]. Fig. 31 shows Neubert's result for the production of transversely and longitudinally polarized $D^*$ mesons in $B \to D^*\ l\ \nu$ decays. Using this figure we find $\Gamma_L/\Gamma$ to be approximately 85% for $q^2 = {m_\rho}^2 = 0.6$, which agrees well with Rosner's prediction [59].

The agreement between these predictions and the experimental result (see section V B)

$$\Gamma_L/\Gamma \ = \ (93 \pm 5 \pm 5)\% \tag{18}$$

supports the factorization hypothesis in hadronic $B$ meson decay for $q^2$ values up to $m_\rho^2$.

## XI. TESTS OF SPIN SYMMETRY IN HQET

If the factorization hypothesis holds, then certain hadronic $B$ meson decay modes can be used to test the spin symmetry of HQET. In this theory the effect of the heavy quark magnetic moment does not enter to lowest order [61] , so it is expected that

$$\Gamma(\bar{B}^0 \to D^+\pi^-) = \Gamma(\bar{B}^0 \to D^{*+}\pi^-) \tag{19}$$

and

$$\Gamma(\bar{B}^0 \to D^+\rho^-) = \Gamma(\bar{B}^0 \to D^{*+}\rho^-). \tag{20}$$

After correcting for phase space and deviations from heavy quark symmetry, one expects that $\mathcal{B}(\bar{B}^0 \to D^+\pi^-) = 1.03\ \mathcal{B}(\bar{B}^0 \to D^{*+}\pi^-)$ and $\mathcal{B}(\bar{B}^0 \to D^+\rho^-) = 0.89\ \mathcal{B}(\bar{B}^0 \to D^{*+}\rho^-)$. A separate calculation by Blok and Shifman using a QCD sum rule approach predicts that $\mathcal{B}(\bar{B}^0 \to D^+\pi^-) = 1.2\ \mathcal{B}(\bar{B}^0 \to D^{*+}\pi^-)$ due to the presence of non-factorizable contributions [62]. From our data we find

$$\frac{\mathcal{B}(\bar{B}^0 \to D^+\pi^-)}{\mathcal{B}(\bar{B}^0 \to D^{*+}\pi^-)} \ = \ 1.12 \pm 0.19 \pm 0.24 \tag{21}$$

and

$$\frac{\mathcal{B}(\bar{B}^0 \to D^+\rho^-)}{\mathcal{B}(\bar{B}^0 \to D^{*+}\rho^-)} \ = \ 1.10 \pm 0.14 \pm 0.28 \tag{22}$$

The contribution in this ratio from the systematic error on the detection efficiency is reduced to 5% for these two cases. Both ratios of branching fractions are consistent with the expectation from HQET spin symmetry as well as the prediction from Blok and Shifman [62].

Mannel *et al.* [61], also observe that by using a combination of HQET, factorization, and data on $B \to D^*\ell\nu$ from CLEO and ARGUS they can obtain model dependent predictions for $\mathcal{B}(\bar{B}^0 \to D^+\rho^-)/\mathcal{B}(\bar{B}^0 \to D^+\pi^-)$. With three different parameterizations of the $B \to D$ form factor [63] this ratio is predicted to be 3.05, 2.52, or 2.61.

From the measurements of the branching ratios we obtain

$$\frac{\mathcal{B}(\bar{B}^0 \to D^+\rho^-)}{\mathcal{B}(\bar{B}^0 \to D^+\pi^-)} \ = \ 2.8 \pm 0.5 \pm 0.2 \tag{23}$$

The systematic errors from the $D$ branching fractions and the tracking efficiency cancel in this ratio. Thus we find good agreement with the prediction from HQET combined with factorization.

## XII. DIRECT AND INDIRECT EFFECTS OF THE COLOR SUPPRESSED AMPLITUDE

### A. Introduction

In the QCD treatment described by equations (1) and (2) it is difficult to take into account the effects of multiple soft gluon emission analytically. Instead, in the phenomenological BSW approach [64] two undetermined coefficients are assigned to the effective charged current, $a_1(\mu)$, and the effective neutral current, $a_2(\mu)$, parts of the $B$ decay Hamiltonian. These coefficients were determined from a fit to a subset of the experimental data on charm decays. With these values the decay rates for a large number of non-leptonic decays can then be calculated using the factorization hypothesis, and model dependent hadron form factors. We can relate $a_1(\mu)$ and $a_2(\mu)$ to the QCD coefficients $c_1(\mu)$ and $c_2(\mu)$ by $a_1 = c_1 + \xi c_2$ and $a_2 = c_2 + \xi c_1$ where $\xi = 1/N_{\rm color}$. The values $a_1(m_c^2) = 1.3$ and $a_2(m_c^2) = -0.55$ which give the best fit to the experimental data on charm decay correspond to $1/N_{\rm color} \sim 0$ [8]. However, there is no rigorous theoretical justification for this choice of $N_{\rm color}$ [65].

In the decays of charmed mesons the effect of color suppression is obscured by the effects of final state interactions (FSI) and soft gluon effects which enhance $W$ exchange diagrams. For instance, Table XVIII gives ratios of several charmed meson decay modes with approximately equal phase space factors where the mode in the numerator is color suppressed while the mode in the denominator is an external spectator decay [67]. These modes are clearly not suppressed. However, the following decay appears to be suppressed.

$$\frac{\mathcal{B}(D^0 \to \bar{K^0}\rho^0)}{\mathcal{B}(D^0 \to K^-\rho^+)} = 0.08 \pm 0.04 \qquad (24)$$

In contrast to the charm sector where the mechanism of color suppression is obscured, one expects to find in $B$ meson decays a simple and consistent pattern of color suppression. Partly, it is expected that color suppression is more effective at the $b$ quark mass scale than the charm quark mass scale due to the evolution of the strong coupling constant $\alpha_s$ to smaller values. Using the BSW model and extrapolating from $q^2 = m_c^2$ to $q^2 = m_b^2$ using the values from charm decays, gives the predictions $a_1(m_b^2) = 1.1$ and $a_2(m_b^2) = -0.24$ for $B$ decays. Another approach using the factorization hypothesis, HQET and model dependent form factors has been suggested by C. Reader and N. Isgur (RI model) [33]. In this approach, $a_1$ and $a_2$ are determined from QCD (with $1/N_{\rm color} = 1/3$) and color suppressed $B$ decays are expected to occur at about 1/1000 the rate of unsuppressed decays. Observation of these decays at a much greater level would indicate the breakdown of the factorization hypothesis. In section VIII we obtained upper limits for color suppressed $B$ decays with a $D^0$ or $D^{*0}$ meson in the final state. In Table XIX these results are compared to prediction of the BSW and the RI model.

In contrast to charm decays, color suppression seems to be operative in hadronic decays of $B$ mesons. The limits on the color suppressed modes with $D^{0(*)}$ and neutral mesons are still above the level expected in the model of Bauer, Stech and Wirbel. However, the limit on $\bar{B}^0 \to D^0\pi^0$ disagrees with Terasaki's prediction [70] that $\mathcal{B}(\bar{B}^0 \to D^0\pi^0) \approx 1.8\ \mathcal{B}(\bar{B}^0 \to D^+\pi^-)$. To date, the only color suppressed $B$ meson decay modes which have been observed are final states which contain charmonium mesons e.g. $B \to \psi K$ and $B \to \psi K^*$ [71].

### B. Determination of $|a_1|$, $|a_2|$ and the relative sign of $(a_2/a_1)$

In the BSW model [8,64] , the branching fractions of the $B^0$ normalization modes are proportional to $a_1^2$ while the branching fractions of the $B \to \psi$ decay modes depend on $a_2^2$ (Table XX [8]). A fit to the branching ratios that we have measured for the modes $\bar{B}^0 \to D^+\pi^-$, $D^+\rho^-$, $D^{*+}\pi^-$ and $D^{*+}\rho^-$ yields

$$|a_1| = 1.15 \pm 0.04 \pm 0.05 \pm 0.09 \tag{25}$$

and a fit to the modes with $\psi$ mesons in the final state gives

$$|a_2| = 0.26 \pm 0.01 \pm 0.01 \pm 0.02 \tag{26}$$

The first systematic error on $|a_1|$ and $|a_2|$ includes the experimental uncertainties from the charm or charmonium branching ratios, tracking efficiency, background shapes and the value of $|V_{cb}|$, but does not include the theoretical uncertainties. There is a second uncertainty due to the $B$ meson production fractions and lifetimes. These are constrained by the value of $(f_+\tau_+/f_0\tau_0)$ determined from the CLEO II [72] measurement of $\mathcal{B}(B^- \to D^{*0}l^-\nu)/\mathcal{B}(\bar{B}^0 \to D^{*+}l^-\nu) = 1.20 \pm 0.20 \pm 0.19$.

The comparison of $B^-$ and $\bar{B}^0$ modes can be used to distinguish between the two possible choices for the sign of $a_2$ relative to $a_1$. The BSW model, Ref. [8] predicts the following ratios:

$$R_1 = \frac{\mathcal{B}(B^- \to D^0\pi^-)}{\mathcal{B}(\bar{B}^0 \to D^+\pi^-)} = (1 + 1.23a_2/a_1)^2 \tag{27}$$

$$R_2 = \frac{\mathcal{B}(B^- \to D^0\rho^-)}{\mathcal{B}(\bar{B}^0 \to D^+\rho^-)} = (1 + 0.66a_2/a_1)^2 \tag{28}$$

The numerical factor which multiplies $a_2/a_1$ is proportional to the ratio of $B \to D^{(*)}$ to $B \to \pi(\rho)$ form factors as well as the ratio of the $\pi(\rho)$ meson to $D$ meson decay constants. We assume $f_D = f_{D^*} = 220$ MeV [69]. Only the $B \to D^*$ form factor and the $\pi(\rho)$ meson decay constant have been measured experimentally.

Similarly, we define

$$R_3 = \frac{\mathcal{B}(B^- \to D^{*0}\pi^-)}{\mathcal{B}(\bar{B}^0 \to D^{*+}\pi^-)} = (1 + 1.29a_2/a_1)^2 \tag{29}$$

$$R_4 = \frac{\mathcal{B}(B^- \to D^{*0}\rho^-)}{\mathcal{B}(\bar{B}^0 \to D^{*+}\rho^-)} \approx (1 + 0.75a_2/a_1)^2 \tag{30}$$

Table XXI shows a comparison between the experimental results and the two allowed solutions in the BSW model. In these ratios, the systematic errors due to detection efficiency are reduced. In the ratios $R_3$ and $R_4$ the $D^0 \to K^-\pi^+$ branching ratio error does not contribute to the systematic error.

It is important to note that the determination of the sign of $a_2/a_1$ depends on assumptions about the relative production of $B^+$ and $B^0$ mesons at the $\Upsilon(4S)$ resonance, $f_+$ and $f_0$, as well as their lifetimes, $\tau_+$ and $\tau_0$. A least squares fit to the above ratios using the CLEO II value for $(f_+\tau_+/f_0\tau_0)$ [72] gives $a_2/a_1 = 0.23 \pm 0.04 \pm 0.04 \pm 0.10$ where we have ignored uncertainties in the theoretical predictions for $R_1$ through $R_4$. The second systematic error is due to the uncertainty in $(f_+\tau_+/f_0\tau_0)$. As this ratio increases, the value of $a_2/a_1$ decreases. The allowed range of $(f_+\tau_+/f_0\tau_0)$ excludes a negative value of $a_2/a_1$. Other uncertainties in the magnitude of $f_D$ and the $B \to \pi$ form factor can change the magnitude of $a_2/a_1$ but not its sign. This result is consistent with the value of $a_2$ determined from the fit to the $B \to \psi$ decay modes. It disagrees with the theoretical extrapolation from data on charmed meson decay in the BSW model [73] which predicts a negative value for $a_2/a_1$.

## XIII. CONCLUSIONS

We have presented new measurements of $B$ branching ratios, resonant substructure and masses. More accurate branching ratios are given for many modes. The modes $B \to D^*\rho^-$ and $B \to D^*a_1^-$ are clearly seen for the first time.

Using a subset of 702 $B$ meson decays reconstructed in channels with good signal to background ratios we have made a precise measurement of the $\bar{B}^0 - B^-$ mass difference of $0.41 \pm 0.25 \pm 0.19$ MeV.

We have carried out an extensive series of tests of the factorization hypothesis including comparisons of rates for $D^{*+}h^-$ (where $h^- = \pi^-, \rho^-$, or $a_1^-$) with rates for $D^{*+}l^-\bar{\nu}$ at $q^2 = M_h^2$, as well as comparisons of the polarizations in $D^{*+}\rho^-$ with $D^{*+}\ell^-\bar{\nu}_\ell$. In all cases the factorization hypothesis is consistent with the data.

We have made improved measurements of branching ratios of two-body decays with a $\psi$, $\psi'$ or $\chi_c$ meson in the final state. The decay $B \to \psi K^*$ is strongly polarized with $\Gamma_L/\Gamma = 0.80 \pm 0.06 \pm 0.08$. Therefore this mode will be useful for measuring CP violation.

A search for color suppressed decays with a charmed meson and light neutral hadron in the final state shows no positive evidence for such processes. The most stringent limit, $\mathcal{B}(\bar{B}^0 \to D^0\pi^0)/\mathcal{B}(\bar{B}^0 \to D^+\pi^-) < 0.09$, is still above the level where these color suppressed $B$ decays are expected in most models.

The observation of $B \to \psi$ modes shows that color suppressed decays are present. Using only exclusive $B \to \psi$ decays we find a value of the BSW parameter $|a_2| = 0.26 \pm 0.01 \pm 0.01 \pm 0.02$. We also report a new value for the BSW parameter $|a_1| = 1.15 \pm 0.04 \pm 0.05 \pm 0.09$. Comparing $B^+$ and $B^0$ decays, we find $a_2/a_1 = 0.23 \pm 0.04 \pm 0.04 \pm 0.10$. We have shown that the sign of $a_2/a_1$ is positive, in contrast to what is found in charm decays.

## ACKNOWLEDGMENTS

We thank N. Cabibbo, Nathan Isgur, W.S. Hou, Volker Rieckert, and J. L. Rosner for useful discussions. We gratefully acknowledge the effort of the CESR staff in providing us with excellent luminosity and running conditions. J.P.A. and P.S.D. thank the PYI program of the NSF, I.P.J.S. thanks the YI program of the NSF, T.E.B. thanks the University of Hawaii, G.E. thanks the Heisenberg Foundation, K.K.G. thanks the SSC Fellowship program of TNRLC, K.K.G., H.N.N., J.D.R., T.S. and H.Y. thank the OJI program of DOE and P.R.

thanks the A.P. Sloan Foundation for support. This work was supported by the National Science Foundation and the U.S. Dept. of Energy.

# TABLES

TABLE I. $D^0$ and $D^+$ decay modes

| $D$ type | Decay Mode | $\mathcal{B}(\%)$ | $\sigma_{m_D}$(MeV) |
|---|---|---|---|
| $D^0$ | $K^-\pi^+$ | 3.91±0.08±0.17 | 8.5 |
| $D^0$ | $K^+\pi^-\pi^0$ | 12.1±1.1 | 13.0 |
| $D^0$ | $K^-\pi^+\pi^+\pi^-$ | 8.0±0.5 | 8.1 |
| $D^+$ | $K^-\pi^+\pi^+$ | 9.1 ± 1.4 | 7.6 |

TABLE II. $D^*$ decay modes used

| $D^*$ Mode | $\mathcal{B}(\%)$ | $\sigma_{m_{D^*}-m_D}$(MeV) |
|---|---|---|
| $D^{*+} \to D^0\pi^+$ | 68.1±1.6 | 0.8 |
| $D^{*0} \to D^0\pi^0$ | 63.6±4.0 | 1.1 |

TABLE III. Branching Ratios (%) for $B^- \to D^0(n\pi)^-$

| $B^-$ Mode | $D$ mode | $\sigma_{\Delta E}$ (MeV) | # of events | $\epsilon$[a] | $\mathcal{B}(\%)$ | $\mathcal{B}$ average (%) |
|---|---|---|---|---|---|---|
| | $K^-\pi^+$ | 22 | 76.3 ± 9.1 | 0.433 | 0.48±0.06 | |
| $D^0\pi^-$ | $K^-\pi^+\pi^0$ | 26 | 134 ± 15 | 0.193 | 0.62±0.07 | 0.55±0.04±0.05 ± 0.02 |
| | $K^-\pi^+\pi^+\pi^-$ | 20 | 94± 11 | 0.222 | 0.57±0.07 | |
| | $K^-\pi^+$ | 18-38 | 80 ± 9 | 0.155 | 1.40 ± 0.18 | |
| $D^0\rho^-$ | $K^-\pi^+\pi^0$ | 22-42 | 42 ± 9 | 0.036 | 1.04±0.23 | 1.35±0.12±0.14 ± 0.04 |
| | $K^-\pi^+\pi^+\pi^-$ | 17-37 | 90.4±12.1 | 0.079 | 1.53±0.20 | |

[a]This efficiency does not include $D$ branching ratios.

TABLE IV. Branching Ratios (%) for $\bar{B}^0 \to D^+(n\pi)^-$

| $\bar{B}^0$ Mode | $D$ Mode | $\sigma_{\Delta E}$ (MeV) | # of events | $\epsilon$[a] | $\mathcal{B}(\%)$ | $\mathcal{B}$ average (%) |
|---|---|---|---|---|---|---|
| $D^+\pi^-$ | $K^-\pi^+\pi^+$ | 20.5 | 80.6±9.8 | 0.32 | 0.29±0.04 | 0.29±0.04±0.03 ± 0.05 |
| $D^+\rho^-$ | $K^-\pi^+\pi^+$ | 18-38 | 78.9±10.7 | 0.12 | 0.81±0.11 | 0.81±0.11±0.12 ± 0.13 |

[a]This efficiency does not include $D$ branching ratios.

TABLE V. Branching Ratios (%) for $B^- \to D^{*0}(n\pi)$

| $B^-$ Mode | $D^0$ Mode | $\sigma_{\Delta E}$ (MeV) | # of events | $\epsilon$[a] | $\mathcal{B}(\%)$ | $\mathcal{B}$ average (%) |
|---|---|---|---|---|---|---|
| | $K^-\pi^+$ | 25 | 13.3±3.8 | 0.16 | 0.36±0.13 | |
| $D^{*0}\pi^-$ | $K^-\pi^+\pi^0$ | 32 | 37.7±6.9 | 0.08 | 0.63±0.12 | 0.52±0.07±0.06 ± 0.04 |
| | $K^-\pi^+\pi^+\pi^-$ | 21 | 20.0±4.9 | 0.08 | 0.52±0.13 | |
| | $K^-\pi^+$ | 21-41 | 25.7±5.4 | 0.064 | 1.74±0.37 | |
| $D^{*0}\rho^-$ | $K^-\pi^+\pi^0$ | 26-46 | 43.8±7.8 | 0.027 | 2.24±0.40 | 1.68±0.21±0.25 ± 0.12 |
| | $K^-\pi^+\pi^+\pi^-$ | 19-39 | 16.9±4.6 | 0.030 | 1.19±0.35 | |
| | $K^-\pi^+$ | 14 | 5.5±2.9 | 0.048 | 0.51±0.26 | |
| $D^{*0}\pi^-\pi^-\pi^+$[b] | $K^-\pi^+\pi^0$ | 22 | 27.7±7.2 | 0.022 | 1.74±0.45 | 0.94±0.20± 0.16 ± 0.06 |
| | $K^-\pi^+\pi^+\pi^-$ | 15 | 15.0±4.5 | 0.025 | 1.26 ±0.37 | |

[a]This efficiency does not include $D$ or $D^*$ branching ratios.
[b]The three pion mass is required to be between 1.0 GeV and 1.6 GeV consistent with an $a_1$ meson. (If this channel is dominated by $a_1^-$, the branching ratio for $D^{*0}a_1^-$ is twice that for $D^{*0}\pi^-\pi^-\pi^+$.)

TABLE VI. Branching Ratios (%) for $\bar{B}^0 \to D^{*+}(n\pi)^-$

| $\bar{B}^0$ Mode | $D^0$ Mode | $\sigma_{\Delta E}$ (MeV) | # of events | $\epsilon$[a] | $\mathcal{B}(\%)$ | $\mathcal{B}$ average (%) |
|---|---|---|---|---|---|---|
| | $K^-\pi^+$ | 25 | 19.4±4.5 | 0.35 | 0.22±0.05 | |
| $D^{*+}\pi^-$ | $K^-\pi^+\pi^0$ | 32 | 31.9±6.4 | 0.14 | 0.30±0.06 | 0.26±0.03±0.04 ± 0.01 |
| | $K^-\pi^+\pi^+\pi^-$ | 21 | 20.5±5.2. | 0.15 | 0.27±0.07 | |
| | $K^-\pi^+$ | 21.5-41.5 | 21.9±5.2 | 0.12 | 0.71±0.17 | |
| $D^{*+}\rho^-$ | $K^-\pi^+\pi^0$ | 23-43 | 39.8±7.2 | 0.048 | 1.08±0.20 | 0.74±0.10±0.14 ± 0.03 |
| | $K^-\pi^+\pi^+\pi^-$ | 20.5-40.5 | 14.6±4.6 | 0.054 | 0.52±0.17 | |
| | $K^-\pi^+$ | 14 | 13.5±3.9 | 0.096 | 0.58±0.17 | |
| $D^{*+}\pi^-\pi^-\pi^+$[b] | $K^-\pi^+\pi^0$ | 22 | 21.7±5.9 | 0.043 | 0.67±0.18 | 0.63±0.10±0.11 ± 0.02 |
| | $K^-\pi^+\pi^+\pi^-$ | 15 | 13.9±4.4 | 0.042 | 0.65±0.19 | |

[a]This efficiency does not include $D$ or $D^*$ branching ratios.
[b]The three pion mass is required to be between 1.0 GeV and 1.6 GeV consistent with an $a_1$ meson. (If this channel dominated by $a_1^-$, the branching ratio for $D^{*+}a_1^-$ is twice that for $D^{*+}\pi^-\pi^-\pi^+$.)

TABLE VII. Branching Ratios (%) for $B \to D^{**}(n\pi)$

| $B$ Mode | $D$ Mode | $\sigma_{\Delta E}$ (MeV) | $\epsilon^{a}$ | # of events | $\mathcal{B}$ average (%) |
|---|---|---|---|---|---|
| $D^0\pi^+\pi^-$ | $K^-\pi^+$ | 17 | 0.19 | $< 10.1$ | $< 0.16$ |
| $D^+\pi^-\pi^-$ | $K^-\pi^+\pi^+$ | 15.5 | 0.11 | $< 10.3$ | $< 0.14$ |
| $D^{**}(2460)\pi^- \to D^+\pi^-\pi^-$ | $K^-\pi^+\pi^+$ | 16 | 0.21 | $< 5.6$ | $< 0.13$ |
| $D^{**}(2460)\pi^- \to D^0\pi^+\pi^-$ | $K^-\pi^+$ | 17 | 0.26 | $< 5.6$ | $< 0.22$ |
| $D^{**}(2460)\rho^- \to D^+\pi^-\pi^-\pi^0$ | $K^-\pi^+\pi^+$ | 16 | 0.08 | $< 6.1$ | $< 0.47$ |
| $D^{**}(2460)\rho^- \to D^0\pi^+\pi^-\pi^0$ | $K^-\pi^+$ | 17 | 0.11 | $< 5.1$ | $< 0.49$ |
| | $K^-\pi^+$ | 16 | 0.161 | | |
| $D^*\pi^-\pi^-$ | $K^-\pi^+\pi^0$ | 23 | 0.061 | $14.1 \pm 5.4$ | $0.19 \pm 0.07 \pm 0.03 \pm 0.01$ |
| | $K^-\pi^+\pi^-\pi^+$ | 19 | 0.075 | | |
| | $K^-\pi^+$ | 16 | 0.161 | | |
| $D^{**}(2420)\pi^- \to D^{*+}\pi^-\pi^-$ | $K^-\pi^+\pi^0$ | 23 | 0.061 | $8.5 \pm 3.8$ | $0.11 \pm 0.05 \pm 0.02 \pm 0.01$ |
| | $K^-\pi^+\pi^-\pi^+$ | 19 | 0.075 | | |
| | $K^-\pi^+$ | 16 | 0.161 | | |
| $D^{**}(2460)\pi^- \to D^{*+}\pi^-\pi^-$ | $K^-\pi^+\pi^0$ | 23 | 0.061 | $3.5 \pm 2.3$ | $< 0.28$ |
| | $K^-\pi^+\pi^-\pi^+$ | 19 | 0.075 | | |
| | $K^-\pi^+$ | 30 | 0.078 | | |
| $D^{**}(2420)\rho^- \to D^{*+}\pi^-\pi^-\pi^0$ | $K^-\pi^+\pi^0$ | 24 | 0.037 | $3.4 \pm 2.1$ | $< 0.14$ |
| | $K^-\pi^+\pi^-\pi^+$ | 27 | 0.042 | | |
| | $K^-\pi^+$ | 30 | 0.078 | | |
| $D^{**}(2460)\rho^- \to D^{*+}\pi^-\pi^-\pi^0$ | $K^-\pi^+\pi^0$ | 24 | 0.037 | $3.2 \pm 2.4$ | $< 0.5$ |
| | $K^-\pi^+\pi^+\pi^-$ | 27 | 0.042 | | |

[a]The efficiencies do not include the branching ratios for $D$, $D^*$ and $D^{**}$. To determine the $B$ decay branching ratios, we assumed $\mathcal{B}(D^{**0}(2420) \to D^{*+}\pi^-)$ and $\mathcal{B}(D^{**0}(2460) \to D^{*+}\pi^-)$ are 67% and 20% respectively. We also assume that $\mathcal{B}(D^{**0}(2460) \to D^+\pi^-)$ and $\mathcal{B}(D^{**+}(2460) \to D^0\pi^+)$ are 30% and 30% respectively.

TABLE VIII. Branching Ratio for $B \to D^{**}(n\pi)$.

| Mode | CLEO II | Bari model [32] | RI model [33] |
|---|---|---|---|
| $D^{**0}(2420)\pi^-$ | $(11 \pm 5 \pm 2 \pm 1) \times 10^{-4}$ | $4 \times 10^{-4}$ | $7.5 \times 10^{-4} - 13 \times 10^{-4}$ |
| $D^{**0}(2460)\pi^-$ $(D^{**0} \to D^{*+}\pi^-)$ | $< 2.8 \times 10^{-3}$ | $6 \times 10^{-4}$ | $5 \times 10^{-4} - 8 \times 10^{-4}$ |
| $D^{**0}(2460)\pi^-$ $(D^{**0} \to D^{+}\pi^-)$ | $< 1.3 \times 10^{-3}$ | $6 \times 10^{-4}$ | $5 \times 10^{-4} - 8 \times 10^{-4}$ |
| $D^{**+}(2460)\pi^-$ $(D^{**+} \to D^{0}\pi^+)$ | $< 2.2 \times 10^{-3}$ | $6 \times 10^{-4}$ | $5 \times 10^{-4} - 8 \times 10^{-4}$ |
| $D^{**0}(2420)\rho^-$ | $< 1.4 \times 10^{-3}$ | $1 \times 10^{-3}$ | $13 \times 10^{-4} - 24 \times 10^{-4}$ |
| $D^{**0}(2460)\rho^-$ | $< 5 \times 10^{-3}$ | $1 \times 10^{-3}$ | $10 \times 10^{-4} - 20 \times 10^{-4}$ |

TABLE IX. Exclusive $B \to c\bar{c}$ Branching Ratios and 90% Confidence Level Upper Limits (%).

| $B$ Mode | $\sigma(\Delta E)$ | # of events | $\epsilon$ [a] | $\mathcal{B}(\%)$ |
|---|---|---|---|---|
| $B^- \to \psi K^-$ | 13 | $58.7 \pm 7.9$ | 0.47 | $0.110 \pm 0.015 \pm 0.009$ |
| $B^0 \to \psi K^0$ | 13 | $10.0 \pm 3.2$ | 0.34 | $0.075 \pm 0.024 \pm 0.008$ |
| $B^0 \to \psi K^{*0}$ | 12 | $29.0 \pm 5.4$ | 0.23 | $0.169 \pm 0.031 \pm 0.018$ |
| $B^- \to \psi K^{*-}$, $K^{*-} \to K^-\pi^0$ | 21 | $6.0 \pm 2.4$ | 0.07 | $0.218 \pm 0.089 \pm 0.026$ |
| $B^- \to \psi K^{*-}$, $K^{*-} \to K_S^0\pi^-$ | 11 | $6.6 \pm 2.7$ | 0.17 | $0.130 \pm 0.058 \pm 0.018$ |
| $B^- \to \psi K^{*-}$ (combined) | | $12.6 \pm 3.6$ | | $0.178 \pm 0.051 \pm 0.023$ |
| | | | | |
| $B^- \to \psi' K^-$ | $9.8, 11$ | $7.0 \pm 2.6$ | $0.36, 0.15$ | $0.061 \pm 0.023 \pm 0.009$ |
| $B^0 \to \psi' K^0$ | $8.4, 10$ | 0 | $0.28, 0.11$ | $< 0.08$ |
| $B^0 \to \psi' K^{*0}$ | $9.7, 10$ | $4.2 \pm 2.3$ | $0.24, 0.091$ | $< 0.19$ |
| $B^- \to \psi' K^{*-}$, $K^{*-} \to K^-\pi^0$ | $18, 17$ | $1 \pm 1$ | $0.077, 0.023$ | $< 0.56$ |
| $B^- \to \psi' K^{*-}$, $K^{*-} \to K_S^0\pi^-$ | $7.9, 9.8$ | $1 \pm 1$ | $0.16, 0.057$ | $< 0.36$ |
| $B^- \to \psi' K^{*-}$ (combined) | | $2 \pm 1.4$ | | $< 0.30$ |
| | | | | |
| $B^- \to \chi_{c1} K^-$ | 18 | $6 \pm 2.4$ | 0.20 | $0.097 \pm 0.040 \pm 0.009$ |
| $B^0 \to \chi_{c1} K^0$ | 16 | $1 \pm 1$ | 0.14 | $< 0.27$ |
| $B^0 \to \chi_{c1} K^{*0}$ | 15 | $1.2 \pm 1.5$ | 0.13 | $< 0.21$ |
| $B^- \to \chi_{c1} K^{*-}$, $K^{*-} \to K^-\pi^0$ | 15 | 0 | 0.033 | $< 0.67$ |
| $B^- \to \chi_{c1} K^{*-}$, $K^{*-} \to K_s\pi^-$ | 17 | 0 | 0.11 | $< 0.30$ |
| $B^- \to \chi_{c1} K^{*-}$, (combined) | | 0 | | $< 0.21$ |

[a] This efficiency does not include the $\psi$,$\psi'$,$\chi_{c1}$,$K^0$,$K^*$ or $K_S^0$ branching ratios. The two sets of values given for the $\psi'$ channels correspond to the two $\psi'$ decay modes $\psi' \to l^+l^-$ and $\psi' \to \psi\pi^+\pi^-$.

TABLE X. Upper limits (90% C.L) on branching fractions for color suppressed $B$ decays.

| Decay Mode | Events | $\epsilon^{e}$ | U. L. (%) at 90% C. L. |
|---|---|---|---|
| $\bar{B}^0 \to D^0\pi^0$ | $< 20.7$ | $0.32, 0.16, 0.18$ | $< 0.048$ |
| $\bar{B}^0 \to D^0\rho^0$ | $< 19.0$ | $0.21, 0.08, 0.12$ | $< 0.055$ |
| $\bar{B}^0 \to D^0\eta$ | $< 9.5$ | $0.31, 0.11, 0.16$ | $< 0.068$ |
| $\bar{B}^0 \to D^0\eta'$ | $< 3.5$ | $0.18, 0.08, 0.11$ | $< 0.086$ |
| $\bar{B}^0 \to D^0\omega$ | $< 12.7$ | $0.16, 0.07, 0.09$ | $< 0.063$ |
| $\bar{B}^0 \to D^{*0}\pi^0$ | $< 11.0$ | $0.13, 0.07, 0.07$ | $< 0.097$ |
| $\bar{B}^0 \to D^{*0}\rho^0$ | $< 8.1$ | $0.09, 0.04, 0.04$ | $< 0.117$ |
| $\bar{B}^0 \to D^{*0}\eta$ | $< 2.3$ | $0.11, 0.05, 0.06$ | $< 0.069$ |
| $\bar{B}^0 \to D^{*0}\eta'$ | $< 2.3$ | $0.07, 0.03, 0.03$ | $< 0.27$ |
| $\bar{B}^0 \to D^{*0}\omega$ | $< 9.0$ | $0.06, 0.03, 0.03$ | $< 0.21$ |

[e]The efficiencies for the $D^0 \to K^-\pi^+$, $D^0 \to K^-\pi^+\pi^0$, and $D^0 \to K^-\pi^+\pi^-\pi^+$ modes are given. These efficiencies do not include $D$, $D^*$ $\eta$, $\eta'$ and $\omega$ branching ratios.

TABLE XI. Upper limits on ratios of branching ratios for color suppressed to normalization modes.

| Ratio of Branching Ratios | U.L. (90% C.L.) |
|---|---|
| $\mathcal{B}(\bar{B}^0 \to D^0\pi^0)/\mathcal{B}(B^- \to D^0\pi^-)$ | $< 0.09$ |
| $\mathcal{B}(\bar{B}^0 \to D^0\rho^0)/\mathcal{B}(B^- \to D^0\rho^-)$ | $< 0.05$ |
| $\mathcal{B}(\bar{B}^0 \to D^0\eta)/\mathcal{B}(B^- \to D^0\pi^-)$ | $< 0.12$ |
| $\mathcal{B}(\bar{B}^0 \to D^0\eta')/\mathcal{B}(B^- \to D^0\pi^-)$ | $< 0.16$ |
| $\mathcal{B}(\bar{B}^0 \to D^0\omega)/\mathcal{B}(B^- \to D^0\rho^-)$ | $< 0.05$ |
| $\mathcal{B}(\bar{B}^0 \to D^{*0}\pi^0)/\mathcal{B}(B^- \to D^{*0}\pi^-)$ | $< 0.20$ |
| $\mathcal{B}(\bar{B}^0 \to D^{*0}\rho^0)/\mathcal{B}(B^- \to D^{*0}\rho^-)$ | $< 0.07$ |
| $\mathcal{B}(\bar{B}^0 \to D^{*0}\eta)/\mathcal{B}(B^- \to D^{*0}\pi^-)$ | $< 0.14$ |
| $\mathcal{B}(\bar{B}^0 \to D^{*0}\eta')/\mathcal{B}(B^- \to D^{*0}\pi^-)$ | $< 0.54$ |
| $\mathcal{B}(\bar{B}^0 \to D^{*0}\omega)/\mathcal{B}(B^- \to D^{*0}\rho^-)$ | $< 0.09$ |

TABLE XII. $B$ Masses from individual modes (not corrected for initial state radiation).

| | $B^-$ Modes | | | $\bar{B}^0$Modes | |
|---|---|---|---|---|---|
| Mode | Mass (MeV) | Events | Mode | Mass (MeV) | Events |
| $D^{*0}\pi^-$ | 5279.7±0.4 | 73 | $D^{*+}\pi^-$ | 5280.1±0.4 | 73 |
| $D^{*0}\rho^-$ | 5280.2±0.4 | 89 | $D^{*+}\rho^-$ | 5280.5±0.4 | 79 |
| $\psi K^-$ | 5279.8±0.4 | 44 | $\psi K^{*0}$ | 5280.4±0.5 | 29 |
| $D^0\pi^-$ | 5279.9±0.3 | 76 | $D^+\pi^-$ | 5280.4±0.3 | 80 |
| $D^0\rho^-$ | 5279.7±0.4 | 80 | $D^+\rho^-$ | 5280.3±0.4 | 79 |
| All | 5279.9±0.2 | 362 | All | 5280.3±0.2 | 340 |

TABLE XIII. Contributions to the systematic error in the $B^0 - B^-$ mass difference.

| | |
|---|---|
| Event sample | 0.15 MeV |
| Background shape | 0.12 MeV |
| $\gamma$ energy calibration | $< 0.03$ MeV |
| Width of $B$ mass peak | $< 0.02$ MeV |
| Track momentum scale | $< 0.01$ MeV |
| total | 0.19 MeV |

TABLE XIV. Measurements of the $\bar{B}^0$–$B^-$ Mass difference.

| Experiment | $M(\bar{B}^0)$–$M(B^-)$ (MeV) |
|---|---|
| CLEO 87 [2] | $2.0 \pm 1.1 \pm 0.3$ |
| ARGUS [6] | $-0.9 \pm 1.2 \pm 0.5$ |
| CLEO 92 [3] | $-0.4 \pm 0.6 \pm 0.5$ |
| CLEO 93 (this result) | $0.41 \pm 0.25 \pm 0.19$ |
| Average | $0.4 \pm 0.3$ |

TABLE XV. Ingredients for Factorization Tests.

| | |
|---|---|
| $\|c_1\|$ | $1.12 \pm 0.10$ |
| $f_\pi$ | $131.74 \pm 0.15$ MeV |
| $f_\rho$ | $215 \pm 4$ MeV |
| $f_{a_1}$ | $205 \pm 16$ MeV |
| $V_{ud}$ | $0.975 \pm 0.001$ |
| $\frac{d\mathcal{B}}{dq^2}(B \to D^* l\ \nu)\|_{q^2=m_\pi^2}(WSB)$ | 0.0023 GeV$^{-2}$ |
| $\frac{d\mathcal{B}}{dq^2}(B \to D^* l\ \nu)\|_{q^2=m_\pi^2}(ISGW)$ | 0.0020 GeV$^{-2}$ |
| $\frac{d\mathcal{B}}{dq^2}(B \to D^* l\ \nu)\|_{q^2=m_\pi^2}(KS)$ | 0.0024 GeV$^{-2}$ |
| $\frac{d\mathcal{B}}{dq^2}(B \to D^* l\ \nu)\|_{q^2=m_\rho^2}(WSB)$ | 0.0025 GeV$^{-2}$ |
| $\frac{d\mathcal{B}}{dq^2}(B \to D^* l\ \nu)\|_{q^2=m_\rho^2}(ISGW)$ | 0.0024 GeV$^{-2}$ |
| $\frac{d\mathcal{B}}{dq^2}(B \to D^* l\ \nu)\|_{q^2=m_\rho^2}(KS)$ | 0.0027 GeV$^{-2}$ |
| $\frac{d\mathcal{B}}{dq^2}(B \to D^* l\ \nu)\|_{q^2=m_{a_1}^2}(WSB)$ | 0.0032 GeV$^{-2}$ |
| $\frac{d\mathcal{B}}{dq^2}(B \to D^* l\ \nu)\|_{q^2=m_{a_1}^2}(ISGW)$ | 0.0030 GeV$^{-2}$ |
| $\frac{d\mathcal{B}}{dq^2}(B \to D^* l\ \nu)\|_{q^2=m_{a_1}^2}(KS)$ | 0.0033 GeV$^{-2}$ |

TABLE XVI. Comparison of $R_{\rm exp}$ and $R_{\rm theor}$

| | R$_{\rm exp}$ (GeV$^2$) | R$_{\rm theor}$ (GeV$^2$) |
|---|---|---|
| $\bar{B}^0 \to D^{*+}\pi^-$ | $1.1 \pm 0.1 \pm 0.2$ | $1.2 \pm 0.2$ |
| $\bar{B}^0 \to D^{*+}\rho^-$ | $3.0 \pm 0.4 \pm 0.6$ | $3.3 \pm 0.5$ |
| $\bar{B}^0 \to D^{*+}a_1^-$ | $4.0 \pm 0.6 \pm 0.5$ | $3.0 \pm 0.5$ |

TABLE XVII. Ratios of $B$ decay widths.

| | Exp. | Factorization | RI Model | BSW Model |
|---|---|---|---|---|
| $\mathcal{B}(\bar{B}^0 \to D^{*+}\rho^-)/\mathcal{B}(\bar{B}^0 \to D^{*+}\pi^-)$ | $2.9 \pm 0.5 \pm 0.5$ | $2.9 \pm 0.05$ | 2.2 – 2.3 | 2.8 |
| $\mathcal{B}(\bar{B}^0 \to D^{*+}a_1^-)/\mathcal{B}(\bar{B}^0 \to D^{*+}\pi^-)$ | $5.0 \pm 1.0 \pm 0.6$ | $3.4 \pm 0.3$ | 2.0 – 2.1 | 3.4 |

TABLE XVIII. Ratios of color suppressed to external spectator branching ratios.

| | |
|---|---|
| $\mathcal{B}(D^0 \to \bar{K}^0\pi^0)/\mathcal{B}(D^0 \to K^-\pi^+)$ | $0.57 \pm 0.13$ |
| $\mathcal{B}(D^0 \to \bar{K}^{*0}\pi^0)/\mathcal{B}(D^0 \to K^{*-}\pi^+)$ | $0.47 \pm 0.23$ |
| $\mathcal{B}(D^0 \to \pi^0\pi^0)/\mathcal{B}(D^0 \to \pi^-\pi^+)$ | $0.77 \pm 0.25$ |
| $\mathcal{B}(D_s^+ \to \bar{K}^{*0}K^+)/\mathcal{B}(D_s \to \phi\pi^+)$ | $0.95 \pm 0.10$ |
| $\mathcal{B}(D_s^+ \to \bar{K}^0K^+)/\mathcal{B}(D_s \to \phi\pi^+)$ | $1.01 \pm 0.16$ |

TABLE XIX. Branching fractions of color suppressed $B$ decays and comparisons with models.

| Decay Mode | U. L. (%) | BSW (%) | $\mathcal{B}$ (BSW) | RI model(%) |
|---|---|---|---|---|
| $\bar{B}^0 \to D^0\pi^0$ | $< 0.048$ | 0.012 | $0.20a_2^2(f_D/220\text{MeV})^2$ | 0.0013 – 0.0018 |
| $\bar{B}^0 \to D^0\rho^0$ | $< 0.055$ | 0.008 | $0.14a_2^2(f_D/220\text{MeV})^2$ | 0.00044 |
| $\bar{B}^0 \to D^0\eta$ | $< 0.068$ | 0.006 | $0.11a_2^2(f_D/220\text{MeV})^2$ | |
| $\bar{B}^0 \to D^0\eta'$ | $< 0.086$ | 0.002 | $0.03a_2^2(f_D/220\text{MeV})^2$ | |
| $\bar{B}^0 \to D^0\omega$ | $< 0.063$ | 0.008 | $0.14a_2^2(f_D/220\text{MeV})^2$ | |
| $\bar{B}^0 \to D^{*0}\pi^0$ | $< 0.097$ | 0.012 | $0.21a_2^2(f_{D*}/220\text{MeV})^2$ | 0.0013 – 0.0018 |
| $\bar{B}^0 \to D^{*0}\rho^0$ | $< 0.117$ | 0.013 | $0.22a_2^2(f_{D*}/220\text{MeV})^2$ | 0.0013 – 0.0018 |
| $\bar{B}^0 \to D^{*0}\eta$ | $< 0.069$ | 0.007 | $0.12a_2^2(f_{D*}/220\text{MeV})^2$ | |
| $\bar{B}^0 \to D^{*0}\eta'$ | $< 0.27$ | 0.002 | $0.03a_2^2(f_{D*}/220\text{MeV})^2$ | |
| $\bar{B}^0 \to D^{*0}\omega$ | $< 0.21$ | 0.013 | $0.22a_2^2(f_{D*}/220\text{MeV})^2$ | |

TABLE XX. Branching ratios in terms of BSW parameters $a_1$, $a_2$

| Mode | B % |
|---|---|
| $\bar{B}^0 \to D^+\pi^-$ | $0.264a_1^2$ |
| $\bar{B}^0 \to D^+\rho^-$ | $0.621a_1^2$ |
| $\bar{B}^0 \to D^{*+}\pi^-$ | $0.254a_1^2$ |
| $\bar{B}^0 \to D^{*+}\rho^-$ | $0.702a_1^2$ |
| $B^- \to D^0\pi^-$ | $0.265[a_1 + 1.230a_2\ (f_D/220)]^2$ |
| $B^- \to D^0\rho^-$ | $0.622[a_1 + 0.662a_2\ (f_D/220)]^2$ |
| $B^- \to D^{*0}\pi^-$ | $0.255[a_1 + 1.292a_2\ (f_{D^*}/220)]^2$ |
| $B^- \to D^{*0}\rho^-$ | $0.703[a_1^2 + 0.635a_2^2(f_{D^*}/220)^2 + 1.487a_1a_2\ (f_{D^*}/220)]$ |
| $B^- \to \psi K^-$ | $1.819a_2^2$ |
| $B^- \to \psi K^{*-}$ | $2.932a_2^2$ |
| $\bar{B}^0 \to \psi\bar{K}^0$ | $1.817a_2^2$ |
| $\bar{B}^0 \to \psi\bar{K}^{*0}$ | $2.927a_2^2$ |

TABLE XXI. Ratios of normalization modes to determine the sign of $a_2/a_1$. The magnitude of $a_2/a_1$ is the value in the BSW model which agrees with our result from $B \to \psi$ modes.

| Ratio | $a_2/a_1 = -0.24$ | $a_2/a_1 = 0.24$ | CLEO II | RI model |
|---|---|---|---|---|
| $R_1$ | 0.50 | 1.68 | $1.89 \pm 0.26 \pm 0.32$ | $1.20 - 1.28$ |
| $R_2$ | 0.71 | 1.34 | $1.67 \pm 0.27 \pm 0.30$ | $1.09 - 1.12$ |
| $R_3$ | 0.48 | 1.72 | $2.00 \pm 0.37 \pm 0.28$ | $1.19 - 1.27$ |
| $R_4$ | 0.41 | 1.85 | $2.27 \pm 0.41 \pm 0.41$ | $1.10 - 1.36$ |

## XIV. APPENDIX

In this appendix, we provide the product of the $B$ and charm branching fractions for the decay modes measured in this paper so that the results can be easily renormalized when the intermediate branching fractions for $D^0, D^+, D^{*+}, D^{*0}$ and $\psi, \psi', \chi_{c1}$ mesons are known more precisely. The results are given in Tables XXII–XXVIII.

TABLE XXII. Product Branching Fractions (%) for $B^- \to D^0(n\pi)^-$ Modes

| $B^-$ Mode | $D$ Mode | # of events | $\epsilon$ | $\mathcal{B}(B^- \to D^0(n\pi)^-)\times$ $\mathcal{B}(D^0 \to K^-[n\pi])$ |
|---|---|---|---|---|
| | $K^-\pi^+$ | 76.3±9.1 | 0.433 | $0.0189 \pm 0.0022 \pm 0.0013$ |
| $D^0\pi^-$ | $K^-\pi^+\pi^0$ | 134±15 | 0.193 | $0.0746 \pm 0.0082 \pm 0.0065$ |
| | $K^-\pi^+\pi^+\pi^-$ | 94± 11 | 0.222 | $0.0455 \pm 0.0055 \pm 0.0049$ |
| | $K^-\pi^+$ | 80 ± 9 | 0.155 | $0.0524 \pm 0.0067 \pm 0.0044$ |
| $D^0\rho^-$ | $K^-\pi^+\pi^0$ | 42 ± 9 | 0.036 | $0.1254 \pm 0.0282 \pm 0.0150$ |
| | $K^-\pi^+\pi^+\pi^-$ | 90.4±12.1 | 0.079 | $0.1223 \pm 0.0164 \pm 0.0142$ |

TABLE XXIII. Product Branching Fractions (%) for $\bar{B}^0 \to D^+(n\pi)^-$ Modes

| $\bar{B}^0$ Mode | $D$ Mode | # of events | $\epsilon$ | $\mathcal{B}(\bar{B}^0 \to D^+(n\pi)^-)\times$ $\mathcal{B}(D^+ \to K^-\pi^+\pi^+)$ |
|---|---|---|---|---|
| $D^+\pi^-$ | $K^-\pi^+\pi^+$ | 80.6±9.8 | 0.32 | $0.0265 \pm 0.0032 \pm 0.0023$ |
| $D^+\rho^-$ | $K^-\pi^+\pi^+$ | 78.9±10.7 | 0.12 | $0.0704 \pm 0.0096 \pm 0.0070$ |

TABLE XXIV. Product Branching Fractions (%) for $B^- \to D^{*0}(n\pi)$ Modes

| $B^-$ Mode | $D^0$ Mode | # of events | $\epsilon$ | $\mathcal{B}(B^- \to D^{*0}(n\pi)^-)\times$ $\mathcal{B}(D^{*0} \to D^0\pi^0) \times \mathcal{B}(D^0 \to K^-[n\pi])$ |
|---|---|---|---|---|
| | $K^-\pi^+$ | 13.3±3.8 | 0.16 | $0.0090 \pm 0.0026 \pm 0.0009$ |
| $D^{*0}\pi^-$ | $K^-\pi^+\pi^0$ | 37.7±6.9 | 0.08 | $0.0488 \pm 0.0089 \pm 0.0063$ |
| | $K^-\pi^+\pi^+\pi^-$ | 20.0±4.9 | 0.08 | $0.0267 \pm 0.0065 \pm 0.0033$ |
| | $K^-\pi^+$ | 25.7±5.4 | 0.064 | $0.0432 \pm 0.0090 \pm 0.0058$ |
| $D^{*0}\rho^-$ | $K^-\pi^+\pi^0$ | 43.8±7.8 | 0.027 | $0.1722 \pm 0.0305 \pm 0.0300$ |
| | $K^-\pi^+\pi^+\pi^-$ | 16.9±4.6 | 0.030 | $0.0608 \pm 0.0176 \pm 0.0095$ |
| | $K^-\pi^+$ | 5.5±2.9 | 0.048 | $0.0124 \pm 0.0065 \pm 0.0020$ |
| $D^{*0}\pi^-\pi^-\pi^+$[a] | $K^-\pi^+\pi^0$ | 27.7±7.2 | 0.022 | $0.1316 \pm 0.0343 \pm 0.0237$ |
| | $K^-\pi^+\pi^+\pi^-$ | 15.0±4.5 | 0.025 | $0.0632 \pm 0.0187 \pm 0.0118$ |

[a]The three pion mass is required to be between 1.0 GeV and 1.6 GeV consistent with an $a_1$ meson. (If this channel is dominated by $a_1^-$, the branching ratio for $D^{*0}a_1^-$ is twice that for $D^{*0}\pi^-\pi^-\pi^+$.)

TABLE XXV. Product Branching Fractions (%) for $\bar{B}^0 \to D^{*+}(n\pi)^-$ Modes

| $\bar{B}^0$ Mode | $D^0$ Mode | # of events | $\epsilon$[a] | $\mathcal{B}(\bar{B}^0 \to D^{*+}(n\pi)^-)\times$ $\mathcal{B}(D^{*+} \to D^0\pi^+) \times \mathcal{B}(D^0 \to K^-[n\pi])$ |
|---|---|---|---|---|
| | $K^-\pi^+$ | 19.4±4.5 | 0.35 | $0.0058 \pm 0.0013 \pm 0.0008$ |
| $D^{*+}\pi^-$ | $K^-\pi^+\pi^0$ | 31.9±6.4 | 0.14 | $0.0243 \pm 0.0049 \pm 0.0035$ |
| | $K^-\pi^+\pi^+\pi^-$ | 20.5±5.2. | 0.15 | $0.0146 \pm 0.0033 \pm 0.0025$ |
| | $K^-\pi^+$ | 21.9±5.2 | 0.12 | $0.0188 \pm 0.0044 \pm 0.0034$ |
| $D^{*+}\rho^-$ | $K^-\pi^+\pi^0$ | 39.8±7.2 | 0.048 | $0.0892 \pm 0.0162 \pm 0.0177$ |
| | $K^-\pi^+\pi^+\pi^-$ | 14.6±4.6 | 0.054 | $0.0286 \pm 0.0091 \pm 0.0059$ |
| | $K^-\pi^+$ | 13.5±3.9 | 0.096 | $0.0151 \pm 0.0044 \pm 0.0024$ |
| $D^{*+}\pi^-\pi^-\pi^+$[a] | $K^-\pi^+\pi^0$ | 21.7±5.9 | 0.043 | $0.0545 \pm 0.0147 \pm 0.0091$ |
| | $K^-\pi^+\pi^+\pi^-$ | 13.9±4.4 | 0.042 | $0.0348 \pm 0.0101 \pm 0.0069$ |

[a]The three pion mass is required to be between 1.0 GeV and 1.6 GeV consistent with an $a_1$ meson. (If this channel dominated by $a_1^-$, the branching ratio for $D^{*+}a_1^-$ is twice that for $D^{*+}\pi^-\pi^-\pi^+$.)

TABLE XXVI. Product Branching Fractions for $B \to \psi$ Modes and 90% Confidence Level Upper Limits (%).

| $B$ Mode | # of events | $\epsilon$ | $\mathcal{B}(B \to \psi K^{(*)}) \times \mathcal{B}(\psi \to l^+l^-)$ [a] |
|---|---|---|---|
| $B^- \to \psi K^-$ | $58.7 \pm 7.9$ | 0.47 | $0.0131 \pm 0.0017 \pm 0.0011$ |
| $B^0 \to \psi K^0$ | $10.0 \pm 3.2$ | 0.34 | $0.0088 \pm 0.0028 \pm 0.0009$ |
| $B^0 \to \psi K^{*0}$ | $29.0 \pm 5.4$ | 0.23 | $0.0200 \pm 0.0037 \pm 0.0021$ |
| $B^- \to \psi K^{*-}$ | $12.6 \pm 3.6$ | | $0.0210 \pm 0.0061 \pm 0.0026$ |

[a]The product branching fraction has been corrected for the $K^0$, $K^*$ or $K_S^0$ branching ratios but not for the $\psi$ branching fractions.

TABLE XXVII. Product Branching Fractions for $B \to \psi'$ Modes and 90% Confidence Level Upper Limits (%).

| $B$ Mode | $\mathcal{B}(B \to \psi' K^{(*)}) \times \mathcal{B}(\psi')$ [a] |
|---|---|
| $B^- \to \psi' K^- \psi' \to l^+ l-$ | $0.0011 \pm 0.0006 \pm 0.0001$ |
| $B^- \to \psi' K^-, \psi' \to \psi \pi^+ \pi^-$ | $0.0173 \pm 0.0100 \pm 0.0023$ |
| $B^0 \to \psi' K^0, \psi' \to l^+ l^-$ | $< 0.0025$ |
| $B^0 \to \psi' K^0, \psi' \to \psi \pi^+ \pi^-$ | $< 0.0200$ |
| $B^0 \to \psi' K^{*0}, \psi' \to l^+ l^-$ | $< 0.0051$ |
| $B^0 \to \psi' K^{*0}, \psi' \to \psi \pi^+ \pi^-$ | $< 0.0210$ |
| $B^- \to \psi' K^{*-}, \psi' \to l^+ l^-$ | $< 0.0065$ |
| $B^- \to \psi' K^{*-}, \psi' \to \psi \pi^+ \pi^-$ | $< 0.0600$ |

[a]The product branching fraction has been corrected for the $K^0$, $K^*$ or $K^0_S$ branching ratios but not for the $\psi'$ and $\psi$ branching fractions. We give $\mathcal{B}(B \to \psi' K^{(*)}) \times \mathcal{B}(\psi' \to l^+ l^-)$ or $\mathcal{B}(B \to \psi' K^{(*)}) \times \mathcal{B}(\psi' \to \psi \pi^+ \pi^-) \times \mathcal{B}(\psi \to l^+ l^-)$.

TABLE XXVIII. Product Branching Fractions for $B \to \chi_{c1}$ Modes and 90% Confidence Level Upper Limits (%).

| $B$ Mode | # of events | $\epsilon$ | $\mathcal{B}(B \to \chi_{c1} K^{(*)}) \times \mathcal{B}(\chi_{c1})$ [a] |
|---|---|---|---|
| $B^- \to \chi_{c1} K^-$ | $6 \pm 2.4$ | 0.20 | $0.0031 \pm 0.0013 \pm 0.0003$ |
| $B^0 \to \chi_{c1} K^0$ | $1 \pm 1$ | 0.14 | $< 0.0087$ |
| $B^0 \to \chi_{c1} K^{*0}$ | $1.2 \pm 1.5$ | 0.13 | $< 0.0066$ |
| $B^- \to \chi_{c1} K^{*-}$ | 0 | | $< 0.0066$ |

[a]For the modes with $\chi_{c1}$ mesons, we report the product $\mathcal{B}(B \to [\chi_{c1}] K^{(*)}) \times \mathcal{B}(\psi \to l^+ l^-) \times \chi_{c1} \to \gamma \psi$ branching fraction in the product. The product branching fraction has been corrected for the $K^0$, $K^*$ or $K^0_S$ branching ratios but not for the $\psi$ branching fractions.

This figure "fig3-1.png" is available in "png" format from:

http://arxiv.org/ps/hep-ph/9403295v1

This figure "fig2-3.png" is available in "png"  format from:

http://arxiv.org/ps/hep-ph/9403295v1

This figure "fig2-6.png" is available in "png" format from:

http://arxiv.org/ps/hep-ph/9403295v1

This figure "fig2-7.png" is available in "png" format from:

http://arxiv.org/ps/hep-ph/9403295v1

This figure "fig2-17.png" is available in "png" format from:

http://arxiv.org/ps/hep-ph/9403295v1

This figure "fig2-22.png" is available in "png" format from:

http://arxiv.org/ps/hep-ph/9403295v1